\documentclass[aps,prd,twocolumn,superscriptaddress,nofootinbib,floatfix,10pt]{revtex4-2}

\usepackage{amsmath,amssymb}
\usepackage{graphicx}
\usepackage{microtype}
\usepackage{xcolor}
\usepackage{etoolbox}

\newif\ifrevisionblue
\revisionbluefalse
\DeclareRobustCommand{\revised}[1]{#1}
\DeclareRobustCommand{\latest}[1]{{\ifrevisionblue\color{blue}\fi #1}}
\newenvironment{revision}{\begingroup}{\endgroup}
\newenvironment{added}{\begingroup\ifrevisionblue\color{blue}\fi}{\endgroup}
\makeatletter
\newcommand{\revisionbibcolor}[1]{%
  \ifrevisionblue
    \ifcsname revisionbib@#1\endcsname\color{blue}\else\color{black}\fi
  \else\color{black}\fi}
\newcommand{\markrevisionbib}[1]{\csdef{#1}{}}
\forcsvlist{\markrevisionbib}{revisionbib@Sanchez:2005}
\newcommand{\revision@bibsetup}{%
  \let\revision@original@lbibitem\@lbibitem
  \def\@lbibitem[##1]##2{%
    \revisionbibcolor{##2}%
    \revision@original@lbibitem[##1]{##2}%
    \revisionbibcolor{##2}\ignorespaces}}
\AtBeginEnvironment{thebibliography}{\revision@bibsetup}
\makeatother
\usepackage[colorlinks=true,linkcolor=blue,citecolor=blue,urlcolor=blue]{hyperref}
\usepackage{orcidlink}
\newcommand{\lz}{l_0}
\newcommand{\Rb}{R_b}
\newcommand{\Mz}{M_0}

\begin{document}

\title{Dressing a naked singularity with zero-point length}

\author{Kimet Jusufi\,\orcidlink{0000-0003-0527-4177}}
\email{kimet.jusufi@unite.edu.mk}
\affiliation{Physics Department, State University of Tetovo, Ilinden Street nn, 1200, Tetovo, North Macedonia}

\author{Francisco S. N. Lobo\,\orcidlink{0000-0002-9388-8373}}
\email{fslobo@ciencias.ulisboa.pt}
\affiliation{Instituto de Astrof\'isica e Ci\^encias do Espa\c{c}o, Faculdade de Ci\^encias da Universidade de Lisboa, Edif\'icio C8, Campo Grande, P-1749-016 Lisbon, Portugal}
\affiliation{Departamento de F\'isica, Faculdade de Ci\^encias da Universidade de Lisboa, Edif\'icio C8, Campo Grande, P-1749-016 Lisbon, Portugal}

\date{\today}

\begin{abstract}
\revised{We regularize the central density cusp of the Joshi--Malafarina--Narayan spacetime using a normalized kernel motivated by a zero-point length. For an unbounded cusp, the convolution and enclosed mass are obtained analytically; truncating this density gives a regular, horizonless tangential-pressure interior matched to Schwarzschild. Convolving the finite cloud instead preserves its total mass and yields a smooth matter tail. A compactness bound establishes horizon avoidance for this second prescription. We determine the central lapse, establish causal geodesic completeness despite extreme optical delays, derive a uniform optical-delay formula across the critical compactness, and identify the stable interior light ring of ultracompact configurations. The massless scalar probe operator is nonnegative, while semiclassical trapped modes exhibit strongly suppressed leakage. An explicit pressure-closure comparison separates robust geometric properties from closure-dependent delay exponents. Planck-scale examples illustrate extreme redshifts and delays, but their photon interpretation requires physics beyond classical propagation in the core. The construction specifies an effective static source, rather than a microscopic matter theory or a dynamical resolution of collapse.}
\end{abstract}

\maketitle

\section{Introduction}

\revised{Classical collapse models can develop spacetime singularities, and the
singularity theorems establish causal geodesic incompleteness under
specified energy, causal, and global hypotheses
\cite{Oppenheimer:1939,Penrose:1965,Hawking:1970}. Weak cosmic censorship
conjectures, in their usual asymptotically flat setting, that singularities
formed by generic collapse from suitable regular initial data are not
visible from future null infinity~\cite{Penrose:1969}. Its validity remains
unsettled. In spherical dust collapse, explicit solutions can produce naked singularities within specified families of regular initial data~\cite{Christodoulou:1984,Joshi:1993,Joshi:2011rev}. These conclusions depend on the matter model and the chosen perturbation class. Scalar-field examples~\cite{Christodoulou:1994} and their instability in the framework of Ref.~\cite{Christodoulou:1999} illustrate why existence is distinct from genericity. Neither result establishes genericity under unrestricted nonspherical perturbations. Whether such objects form in nature is therefore a question
for observation as well as for theory.}

Among the known examples, the first Joshi--Malafarina--Narayan (JMN-1)
spacetime~\cite{JMN:2011} stands out for its simplicity and for its
astrophysical use. It is the static configuration approached asymptotically
by a collapsing cloud with vanishing radial and non-vanishing tangential
pressure. It has no horizon, it is matched smoothly to a Schwarzschild
exterior, and it has a curvature singularity at the centre. Its accretion-disk
spectra and its shadow have been worked out in
detail~\cite{JMN:2014,Shaikh:2019}. \revised{For $\Mz\geq2/3$, its Schwarzschild photon-sphere scale and the images
computed under the accretion assumptions of Ref.~\cite{Shaikh:2019}
can match those of a Schwarzschild black hole. The EHT metric tests
in Ref.~\cite{EHT:2022} do not exclude that singular JMN-1 model.
This statement does not automatically extend to a transparent regular
interior, whose returning radiation requires a separate transport model.} JMN-1 has thus become
one of the standard horizonless alternatives against which the black-hole
paradigm is tested~\cite{Cardoso:2019}.

A naked singularity is a place where the classical theory fails in full view
of distant observers. One expects quantum gravity to intervene there. A
robust expectation, shared by most approaches, is that spacetime has a
zero-point length $\lz$ below which distances lose operational
meaning~\cite{Garay:1995,Hossenfelder:2013}. Padmanabhan showed that such a
length follows from a duality principle for the path integral, which replaces
the squared geodesic interval $\sigma^{2}$ by $\sigma^{2}+\lz^{2}$ in the
propagator~\cite{Padmanabhan:1997,Padmanabhan:1998}. A corresponding effective propagator is motivated by string T-duality,
with $\lz=2\pi\sqrt{\alpha'}$ in the cited convention
\cite{Smailagic:2003,Spallucci:2005,Fontanini:2006,Nicolini:2019}. Applied to a
point mass, it yields a neutral and charged regular black hole with a Bardeen-type
core~\cite{Nicolini:2019,Gaete:2022ukm}, in the same spirit as
noncommutative-geometry-inspired sources~\cite{Nicolini:2006}. The
zero-point length has been used in the context of black hole solutions with different topology and cosmological
settings~\cite{Jusufi:2022rbt,Jusufi:2023b}, and it has recently been argued to
protect regular black holes against transitions to naked
singularities~\cite{Anand:2026}.

The point-source regular-black-hole constructions provide the starting motivation; related cosmological applications involve different settings.
Here we ask a different question: what does a zero-point length do to a
singularity that is already naked and that is produced by an extended matter
distribution? The question is natural for two reasons. First, the JMN-1
singularity is not a delta-function source but an integrable density cusp, so
the way the zero-point length acts on it is not fixed by the point-mass
result.
Second, this is a horizonless configuration. The strong-energy-condition
violation found in familiar regular black holes with de Sitter-like cores
\cite{Bardeen:1968,Hayward:2006,Ansoldi:2008} need not occur here. General
singularity theorems require additional causal and global hypotheses;
regularity alone is not a universal criterion for violating that condition.

Regular static interiors with zero radial pressure and nonzero tangential
stress have classical precedents, notably the Florides construction
\cite{Florides:1974}. Stable light rings in regular horizonless
ultracompact objects are also an established phenomenon
\cite{Cunha:2017}. The contribution here is the explicit convolution-derived
density, its consistent finite-boundary relativistic realization, and the analytic and numerical determination of its lapse and optical delay. We extend these results with a mass-preserving finite-cloud construction, a uniform delay crossover, a nonnegative scalar probe operator, and an explicit comparison of pressure closures. Stable trapping is examined through leading semiclassical mode estimates rather than identified with a proven instability of the effective fluid.

The central questions are whether source smoothing gives a globally regular relativistic geometry, which conclusions survive a consistent finite-source prescription, and how the regular core changes propagation. Section~\ref{sec:jmn} introduces JMN-1 and the kernel, distinguishes the two finite-source prescriptions, and proves a compactness bound for the fully convolved cloud. Section~\ref{sec:geometry} constructs the sharply matched geometry and its redshift. Section~\ref{sec:geodesics} treats orbits, light rings, the uniform delay crossover, infaller proper time and causal completeness, and illustrative mass scales. Section~\ref{sec:probe} proves scalar probe mode stability, estimates trapped modes, and compares pressure closures. Section~\ref{sec:discussion} discusses the contribution and its physical limits. We use geometrized units $G=c=1$ and signature $(-,+,+,+)$, restoring Newton's constant $G$, the speed of light $c$, and Planck's constant where numerical or quantum scales are needed. Length and time consequently have the same dimension, while density and pressure have dimension length$^{-2}$.

\section{\revised{JMN-1 and the regularized source}}
\label{sec:jmn}

\subsection{Metric and source}

Consider a static, spherically symmetric metric
\begin{equation}
ds^{2}=-e^{2\Phi(r)}dt^{2}+\frac{dr^{2}}{1-2m(r)/r}+r^{2}d\Omega^{2},
\label{eq:metric}
\end{equation}
sourced by an anisotropic fluid with density $\rho$, radial pressure $p_{r}$,
and tangential pressure $p_{\theta}$~\cite{Bowers:1974}. The function $m(r)$
is the Misner--Sharp mass~\cite{Misner:1964}. \revised{Here $r\geq0$ is the areal radius, $t$ is normalized at infinity after exterior matching, $\Phi$ is dimensionless, and $d\Omega^2=d\vartheta^2+\sin^2\vartheta\,d\varphi^2$ is the unit-sphere metric. The angular coordinates are $\vartheta$ and $\varphi$; $p_\varphi=p_\theta$ by spherical symmetry. A prime denotes differentiation with respect to the displayed radial argument.} The Einstein equations and the
conservation of the stress tensor give
\begin{align}
m' &= 4\pi r^{2}\rho, \label{eq:ee1}\\
\Phi' &= \frac{m+4\pi r^{3}p_{r}}{r\,(r-2m)}, \label{eq:ee2}\\
p_{r}' &= -(\rho+p_{r})\,\Phi'+\frac{2\,(p_{\theta}-p_{r})}{r}. \label{eq:ee3}
\end{align}

The JMN-1 solution~\cite{JMN:2011} is the member of this family with
$p_{r}=0$ and a linear mass function,
\begin{equation}
\rho=\frac{\Mz}{8\pi r^{2}},\quad p_{r}=0,\quad
p_{\theta}=\frac{a}{2}\,\rho,\quad m(r)=\frac{\Mz\,r}{2},
\label{eq:source}
\end{equation}
where
\begin{equation}
a\equiv\frac{\Mz}{2(1-\Mz)},\qquad 0<\Mz<1.
\end{equation}
Integrating Eq.~\eqref{eq:ee2} gives the interior metric
\begin{equation}
ds^{2}=-(1-\Mz)\left(\frac{r}{\Rb}\right)^{2a}dt^{2}
+\frac{dr^{2}}{1-\Mz}+r^{2}d\Omega^{2}.
\label{eq:jmn}
\end{equation}
It is matched at $r=\Rb$ to a Schwarzschild exterior of mass
\begin{equation}
M=\frac{\Mz\Rb}{2}.
\end{equation}
\revised{With the lapse normalized to Schwarzschild at the boundary, continuity
of $m$ and vanishing $p_r$ ensure continuity of the induced metric and
extrinsic curvature. The Darmois--Israel matching therefore requires no
thin shell~\cite{Israel:1966}. Matter and curvature may have finite jumps;
``shell-free'' does not mean that every metric derivative is continuous
in the areal-radius coordinates.}
The stress tensor \eqref{eq:source} describes a tangential-pressure fluid.
Its interpretation as an Einstein cluster of collisionless massive
particles on circular geodesics~\cite{Einstein:1939} is valid only when
the local circular speed is subluminal, requiring $\Mz<2/3$ in the
singular interior. The general tangential-pressure-fluid construction
\cite{JMN:2011} is broader than this kinetic realization. In the collapse picture of
Ref.~\cite{JMN:2011}, the cloud starts from regular initial data and
approaches Eq.~\eqref{eq:jmn} as the comoving time goes to infinity.

The energy conditions are simple. Since $\rho>0$ and $p_{\theta}>0$, the
weak and strong conditions hold for all $\Mz$. The dominant condition,
$p_{\theta}\leq\rho$, requires $a\leq 2$, that is, $\Mz\leq 4/5$.

\subsection{The singularity}

Since $2m/r=\Mz<1$ there is no trapped surface and no horizon. The density
diverges at the centre, and so does the Kretschmann scalar,
\begin{equation}
R_{\mu\nu\rho\sigma}R^{\mu\nu\rho\sigma}
=\frac{\Mz^{2}\left(a^{2}-2a+7\right)}{r^{4}}.
\label{eq:Kjmn}
\end{equation}
Radial null geodesics obey $dr/d\lambda\propto r^{-a}$, so they reach $r=0$
at finite affine parameter $\lambda\propto r^{a+1}$. The coordinate time
they take is
\begin{equation}
t=\frac{1}{1-\Mz}\int\left(\frac{\Rb}{r}\right)^{a}dr,
\end{equation}
\revised{which is finite for $a<1$ and divergent for $a\geq1$. The same integral
defines the interior tortoise coordinate,
$dr_*/dr=(\Rb/r)^a/(1-\Mz)$, for which the radial metric is conformal
to $-dt^2+dr_*^2$. Its central endpoint is at finite $r_*$ when $a<1$
and at $r_*\to-\infty$ when $a\geq1$. This gives a timelike central
singular boundary for $\Mz<2/3$ and a null one for $\Mz\geq2/3$
\cite{Patel:2026}; coordinate-time divergence alone is not evidence
for an event horizon.} In both cases outgoing null rays from the
vicinity of the centre reach infinity, with a redshift that diverges as the
emission point approaches $r=0$.

\begin{revision}
The density in Eq.~\eqref{eq:source} is locally integrable in the mass
measure, and $m(r)\to0$ at the centre. The tangential pressure nevertheless
diverges with the density. Integrability is not a criterion for weak
curvature strength~\cite{Clarke:1985}. To see this directly, let
$E=-g_{tt}\,dt/d\lambda>0$ and choose the affine endpoint at $r=0$ to
be $\lambda=0$. Along an outgoing radial null geodesic,
\begin{equation}
\frac{dr}{d\lambda}=E\left(\frac{\Rb}{r}\right)^a,
\qquad
\lambda=\frac{r^{a+1}}{(a+1)E\Rb^a}.
\end{equation}
Einstein's equations then give
\begin{equation}
\begin{split}
\lambda^2R_{\mu\nu}k^\mu k^\nu
&=\frac{\Mz}{(1-\Mz)(a+1)^2}\\
&=\frac{4\Mz(1-\Mz)}{(2-\Mz)^2}>0,
\end{split}
\label{eq:nullstrength}
\end{equation}
where $k^\mu=dx^\mu/d\lambda$. The radial null singularity satisfies
the Ricci-focusing sufficient criterion for Tipler-strong curvature
\cite{Clarke:1985}. The motivation for smearing is therefore the
integrability of the density cusp, rather than any assertion that the
original singularity is weak. Full regularity will follow only after
specifying and checking both metric functions and the stresses.
\end{revision}

\subsection{\revised{Zero-point length and the static kernel}}
\label{sec:zpl}

\subsubsection{Modified propagator and static potential}

\revised{Padmanabhan's path-integral duality assigns equal weights to path lengths
$s$ and $L_{\rm d}^{2}/s$, where $L_{\rm d}$ denotes the duality scale
\cite{Padmanabhan:1997,Padmanabhan:1998}. We distinguish this scale from
the interval regulator $\lz$: in the normalization adopted in
Ref.~\cite{Nicolini:2019}, $\lz=2L_{\rm d}$. The massless Euclidean
propagator used here is}
\begin{equation}
G(k)=-\frac{\lz}{\sqrt{k^{2}}}\,K_{1}\!\left(\lz\sqrt{k^{2}}\right),
\label{eq:prop}
\end{equation}
\revised{Here $k^2$ is the squared Euclidean four-momentum and $K_1$ is a modified Bessel function of the second kind. In the static limit the frequency is zero; below, $k=|\mathbf k|$ denotes the spatial Fourier wavenumber. Both momentum magnitudes have dimension inverse length. We use $\sigma^2$ for the squared geodesic interval, rather than for Synge's half-squared world function. The string parameter $\alpha'$ has dimension length squared.} For
$\lz\to0$ one recovers $G=-1/k^{2}$. In position space Eq.~\eqref{eq:prop}
\revised{corresponds, in flat Euclidean space, to the interval replacement
$\sigma^2\to\sigma^2+\lz^2$. In the cited compactified closed-string
construction, a suitable expansion of the T-dual centre-of-mass
propagator yields this effective form, with
$\lz=2\pi\sqrt{\alpha'}$~\cite{Smailagic:2003,Spallucci:2005,Fontanini:2006,Nicolini:2019}.
We use that effective propagator as motivation for a static kernel,
without identifying it with a complete string-theory gravitational action.}

The static potential generated by a point mass $M$ follows from
Eq.~\eqref{eq:prop}~\cite{Nicolini:2019},
\begin{equation}
\phi(r)=-\frac{M}{\sqrt{r^{2}+\lz^{2}}}.
\label{eq:pot}
\end{equation}
Through the Poisson equation, $\nabla^{2}\phi=4\pi\rho$, this is the
potential of a smooth density that replaces the Dirac delta,
\begin{equation}
\delta^{3}(\mathbf{x})\;\to\;K(\mathbf{x})
=\frac{3\,\lz^{2}}{4\pi\left(|\mathbf{x}|^{2}+\lz^{2}\right)^{5/2}},
\label{eq:kernel}
\end{equation}
which is normalized to unity. Its Fourier transform is
$\tilde K(k)=k\lz K_{1}(k\lz)$.

\subsubsection{Smearing an extended source}

The modification acts on the Green function, so it is linear in the source.
\revised{For a static distribution for which the integral converges, the potential is}
\begin{equation}
\phi(\mathbf{x})=-\int d^{3}x'\,
\frac{\rho(\mathbf{x}')}{\sqrt{|\mathbf{x}-\mathbf{x}'|^{2}+\lz^{2}}},
\end{equation}
\revised{When the potential is defined with the corresponding boundary conditions,
this is the ordinary Newtonian potential of the smeared density}
\begin{equation}
\rho_{\rm eff}(\mathbf{x})=\int d^{3}x'\,\rho(\mathbf{x}')\,
K(\mathbf{x}-\mathbf{x}').
\label{eq:conv}
\end{equation}

Following the effective-source approach of
Refs.~\cite{Nicolini:2006,Nicolini:2019}, we prescribe $\rho_{\rm eff}$ as
the density in the nonlinear Einstein equations. This is a phenomenological
promotion of a flat-space static convolution, not a derivation of a
covariant nonlinear gravitational action or of a complete modified stress
tensor. For a source with finite integrated mass, kernel normalization
preserves $\int\rho\,d^3x$ when the entire smeared distribution, including
its exterior tail, is retained. Its interpretation as gravitational mass
also requires specifying the relativistic geometry and pressure closure.

For the exact untruncated profile $\rho=A/r^2$, the central density is
\begin{equation}
\rho_{\rm eff}(0)=4\pi A\int_0^\infty K(s)\,ds.
\label{eq:rho0general}
\end{equation}
A bounded kernel with an integrable radial restriction therefore removes
this cusp. For a source equal to $A/r^2$ only near the centre, regularity
persists provided its outer contribution to the convolution is finite,
but Eq.~\eqref{eq:rho0general} is no longer its exact central density.

\subsection{Closed-form density and mass function}

For the untruncated profile $\rho=A/r^2$ with $A=\Mz/(8\pi)$, the
convolution \eqref{eq:conv} has a closed form. \revised{This infinite profile is an auxiliary local model, not a finite-mass cloud.
Its potential diverges logarithmically in the infrared because its distant
shells contribute $-4\pi A\int^\infty ds/s$. A potential calculation
would therefore require an infrared cutoff or an additive subtraction.
The density convolution itself converges and is the object evaluated below;
no unregulated infinite-cloud potential is inserted into the Einstein
equations. The Fourier transform of the cusp, understood distributionally,
is}
\begin{equation}
\tilde\rho(k)=4\pi A\int_{0}^{\infty}\frac{\sin kr}{kr}\,dr
=\frac{2\pi^{2}A}{k},
\end{equation}
and the convolution multiplies it by $\tilde K(k)=k\lz K_{1}(k\lz)$.
Inverting,
\begin{align}
\rho_{\rm eff}(r)
&=\frac{1}{2\pi^{2}r}\int_{0}^{\infty}k\sin(kr)\,
\tilde\rho(k)\tilde K(k)\,dk \nonumber\\
&=\frac{A\lz}{r}\int_{0}^{\infty}k\sin(kr)\,K_{1}(k\lz)\,dk.
\label{eq:inv}
\end{align}
The remaining integral follows from the standard
result~\cite{Gradshteyn}
\begin{equation}
\int_{0}^{\infty}\sin(kr)\,K_{0}(k\lz)\,dk
=\frac{\sinh^{-1}(r/\lz)}{\sqrt{r^{2}+\lz^{2}}}
\end{equation}
by differentiation with respect to $\lz$, using $K_{0}'=-K_{1}$:
\begin{equation}
\int_{0}^{\infty}k\sin(kr)\,K_{1}(k\lz)\,dk
=\frac{r}{\lz\,(r^{2}+\lz^{2})}+\frac{\lz\sinh^{-1}(r/\lz)}{(r^{2}+\lz^{2})^{3/2}}.
\end{equation}
Inserting this in Eq.~\eqref{eq:inv} gives
\begin{equation}
\rho(r)=\frac{\Mz}{8\pi}\left[\frac{1}{r^{2}+\lz^{2}}
+\frac{\lz^{2}\,\sinh^{-1}(r/\lz)}{r\left(r^{2}+\lz^{2}\right)^{3/2}}\right],
\label{eq:rho}
\end{equation}
where from now on we drop the subscript on $\rho_{\rm eff}$. The mass
function follows from Eq.~\eqref{eq:ee1}. The first term of
Eq.~\eqref{eq:rho} integrates to $r-\lz\arctan(r/\lz)$. The second is
integrated by parts,
\begin{align}
\int_{0}^{r}\frac{\lz^{2}s\sinh^{-1}(s/\lz)}{(s^{2}+\lz^{2})^{3/2}}\,ds
&=-\frac{\lz^{2}\sinh^{-1}(r/\lz)}{\sqrt{r^{2}+\lz^{2}}}\nonumber\\
&\quad+\lz\arctan\frac{r}{\lz},
\end{align}
and the arctangents cancel:
\begin{equation}
m(r)=\frac{\Mz\,r}{2}\,\bigl[1-h(r/\lz)\bigr],\qquad
h(x)\equiv\frac{\sinh^{-1}x}{x\sqrt{1+x^{2}}}.
\label{eq:mass}
\end{equation}
We have checked Eqs.~\eqref{eq:rho} and \eqref{eq:mass} against direct
numerical evaluation of the convolution \eqref{eq:conv}. This check uses
the untruncated source; it does not establish equivalence to a convolved
finite cloud. The function $h$ decreases monotonically from $h(0)=1$ to zero, with
\begin{align}
h(x)&=1-\tfrac{2}{3}x^{2}+\tfrac{8}{15}x^{4}+O(x^{6}),
& x&\ll1, \label{eq:hsmall}
\end{align}
and 
\begin{align}
h(x)&=\frac{\ln(2x)}{x^{2}}+O\!\left(\frac{\ln x}{x^{4}}\right),
& x&\gg1. \label{eq:hlarge}
\end{align}
For $r\gg\lz$ the JMN-1 profile is recovered,
\begin{equation}
\rho\simeq\frac{\Mz}{8\pi r^{2}}
\left[1+\frac{\lz^{2}}{r^{2}}\left(\ln\frac{2r}{\lz}-1\right)\right],
\end{equation}
and the mass function differs from $\Mz r/2$ by a relative correction of
order $(\lz/r)^{2}\ln(r/\lz)$.
We adopt Eq.~\eqref{eq:rho} for $0\le r\le\Rb$ and set the density to
zero outside, with $\Rb\gg\lz$ for the asymptotic results below. This
truncation defines the finite object studied here. It is an additional
boundary prescription, not exact convolution of the original finite cloud.
For a finite bare source, the exact convolution instead reads
\begin{align}
\rho_{\Rb}(r)&=\frac{A\lz^2}{2r}\int_0^{\Rb}\frac{ds}{s}
\biggl\{[(r-s)^2+\lz^2]^{-3/2}\nonumber\\
&\hspace{30mm}-[(r+s)^2+\lz^2]^{-3/2}\biggr\}.
\label{eq:finiteconv}
\end{align}
It is nonzero outside $\Rb$ and has central value
\begin{equation}
\rho_{\Rb}(0)=\frac{A}{\lz^2}(3u-u^3),\qquad
u=\frac{\Rb}{\sqrt{\Rb^2+\lz^2}}.
\end{equation}
For fixed $r$ well inside the boundary, Eq.~\eqref{eq:rho} is recovered
as $\Rb/\lz\to\infty$. This approximation is not uniform near $\Rb$:
$\rho_{\Rb}(\Rb)/(A/\Rb^2)\to1/2$. \revised{Retaining this tail defines a second, mass-preserving object. Its geometry is constructed below; it is distinguished from the sharply truncated model used for the Schwarzschild-matched optical examples.}

\subsection{\revised{The fully convolved finite cloud}}
\begin{figure*}[tbp]
\includegraphics[width=\textwidth]{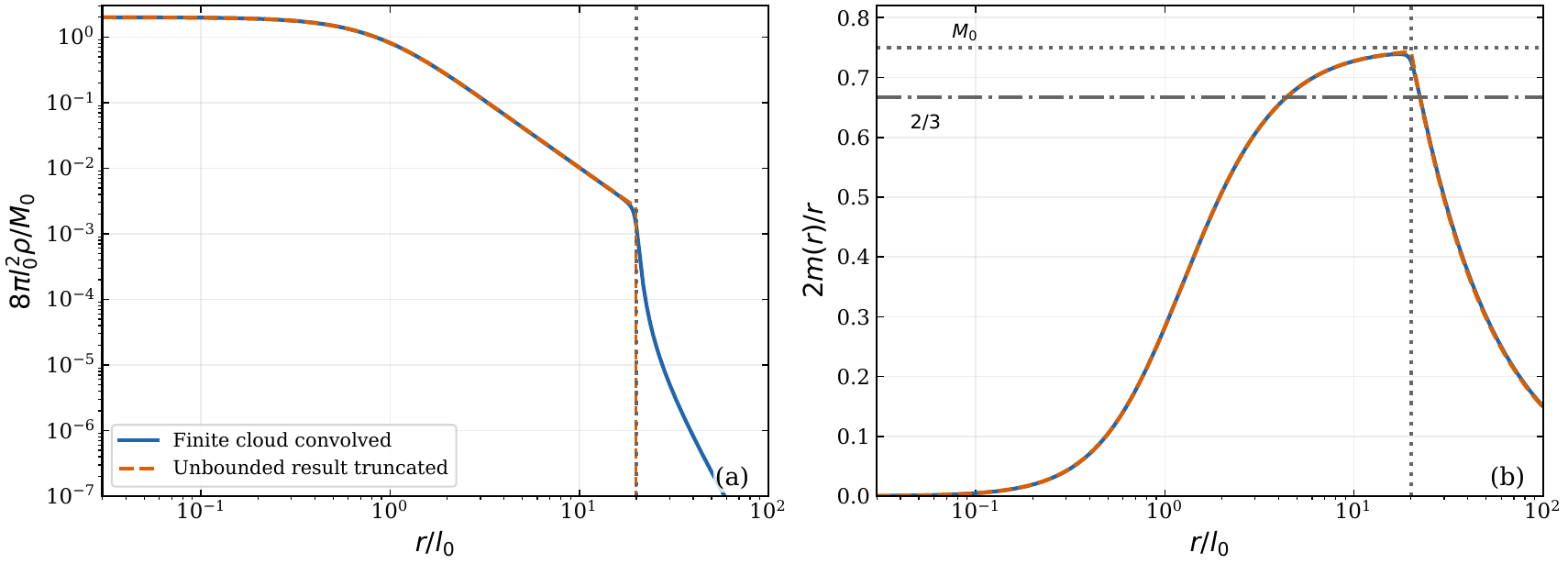}
\caption{\revised{Comparison of finite-source prescriptions for $M_0=0.75$ and $R_b/l_0=20$. Panel (a) plots $8\pi l_0^2\rho/M_0$ against $r/l_0$: convolution of the bare finite cusp with its tail retained (solid) is smooth through $R_b$, while truncation of the unbounded-profile convolution (dashed) produces a finite density jump there. The vertical dotted line marks the bare-source cutoff, which is a material boundary only for the dashed prescription. Panel (b) plots $2m(r)/r$ for both models; the horizontal lines mark the light-ring value $2/3$ and the upper bound $M_0$. The fully convolved total mass is exactly $M_b=M_0R_b/2$, whereas the dashed model has vacuum exterior mass $M=M_0R_b[1-h(R_b/l_0)]/2$. The moderate radius ratio resolves the boundary layer and is not an astrophysical parameter choice.}}
\label{fig:finite}
\end{figure*}

\begin{revision}
Let $\rho_b(r)=A r^{-2}\mathbf1_{r<R_b}$ denote the bare finite source, where $\mathbf1$ is the indicator function and $A=M_0/(8\pi)$. Write $\rho_b*K=\rho_{R_b}$ for Eq.~\eqref{eq:finiteconv}, and define
\begin{equation}
\begin{aligned}
m_{R_b}(r)&=4\pi\int_0^r s^2\rho_{R_b}(s)\,ds,\\
M_b&=4\pi A R_b=\frac{M_0R_b}{2}.
\end{aligned}
\label{eq:finiteM}
\end{equation}
$M_b$ is the total gravitational mass of the fully convolved model. It differs from the Schwarzschild mass $M=m(R_b)$ of the truncated unbounded-profile prescription. Kernel normalization and positivity allow interchange of the integrals, giving $m_{R_b}(\infty)=M_b$ exactly.

\textit{Finite-source compactness bound.} For the present kernel,
\begin{equation}
 m_{R_b}(r)\leq\frac{M_0}{2}\min(r,R_b),
 \qquad \frac{2m_{R_b}(r)}r\leq M_0<1.
\label{eq:finitebound}
\end{equation}
To prove this, integrate the convolution over the Euclidean ball $B_r$ of radius $r$:
\begin{equation}
 m_{R_b}(r)=\int d^3y\,K(\mathbf y)
              \int_{B_r-\mathbf y}\rho_b(\mathbf z)\,d^3z.
\end{equation}
A radial nonincreasing density has its largest integral over a translated ball when the ball is centred at the origin. Indeed, expressing the density as an integral of indicators of concentric superlevel balls reduces the claim to the fact that the intersection volume of two balls is maximal at zero separation. Thus the inner integral is bounded by $\int_{B_r}\rho_b\,d^3z=4\pi A\min(r,R_b)$. Averaging with the normalized nonnegative kernel proves Eq.~\eqref{eq:finitebound}. The mass bound applies to any nonnegative normalized convolution kernel; spherical symmetry of the resulting geometry additionally requires a radial kernel. Smoothness at the centre uses the smooth bounded kernel in Eq.~\eqref{eq:kernel}, not normalization alone.

Impose $p_r=0$ also for this second model. Its exact lapse, normalized at infinity, and tangential pressure are
\begin{align}
 \Phi_{R_b}(r)&=-\int_r^\infty
 \frac{m_{R_b}(s)}{s[s-2m_{R_b}(s)]}\,ds,\label{eq:finitelapse}\\
 p_{\theta,R_b}(r)&=\frac{\rho_{R_b}(r)m_{R_b}(r)}{2[r-2m_{R_b}(r)]}.
\end{align}
The integrand is $O(s)$ at the regular centre and $O(s^{-2})$ at infinity. Equation~\eqref{eq:finitebound} keeps its denominator positive. Hence $e^{2\Phi_{R_b}}$ is strictly positive, the centre is smooth, and the static spacetime is asymptotically flat and horizonless. For every fixed $l_0>0$, its causal geodesics extend through the centre and to arbitrarily large radius without a finite-affine-parameter boundary. The weak and strong energy conditions hold. The dominant condition holds exactly when $\sup_r[2m_{R_b}(r)/r]\leq4/5$; $M_0\leq4/5$ is sufficient by Eq.~\eqref{eq:finitebound}.

For numerical evaluation, an independent enclosed-mass formula follows by angular integration of the auxiliary Newtonian potential and use of its Poisson equation:
\begin{equation}
 m_{R_b}(r)=2\pi A\int_0^{R_b}\frac{ds}{s}
 \biggl[d_+-d_--r\left(\frac{r+s}{d_+}-\frac{r-s}{d_-}\right)\biggr]
 \label{eq:finiteGauss},
\end{equation}
where $d_\pm=\sqrt{(r\pm s)^2+l_0^2}$. This is a formula for the same $4\pi\int\rho_{R_b}r^2dr$ used in Einstein's equations; it does not identify the relativistic lapse with the Newtonian potential. At $r\gg R_b,l_0$ the retained tail satisfies
\begin{align}
 \rho_{R_b}(r)&=\frac{3l_0^2M_b}{4\pi r^5}
 \left[1+O\left(\frac{R_b^2+l_0^2}{r^2}\right)\right],\\
 m_{R_b}(r)&=M_b-\frac{3l_0^2M_b}{2r^2}
 +O\left(\frac{l_0^2M_b(R_b^2+l_0^2)}{r^4}\right).
 \label{eq:finitetail}
\end{align}
Thus there is no exact vacuum matching surface, even though the far field approaches Schwarzschild. Here $R_b$ is the cutoff radius of the bare source, rather than the edge of the convolved matter. Figure~\ref{fig:finite} compares both prescriptions for a display-scale core. In this second model circular null orbits still satisfy $2m_{R_b}/r=2/3$. Since this compactness vanishes at both endpoints $r=0$ and $r=\infty$, a value above $2/3$ implies at least two crossings; simple upward and downward crossings yield stable and unstable light rings, respectively. Their locations need not equal the Schwarzschild-matched values. For the displayed $M_0=0.75$, $R_b/l_0=20$ example, numerical evaluation gives interior and exterior ring radii $r/l_0\simeq4.3762$ and $22.3187$, respectively. The asymptotic-mass Schwarzschild value would be $3M_b/l_0=22.5$, demonstrating the effect of the retained tail. Defining the finite-cloud radial time between the two bare-source cutoff radii by $T_b=2\int_0^{R_b}e^{-\Phi_{R_b}}[1-2m_{R_b}/r]^{-1/2}dr$ gives $T_b/R_b\simeq29.307$, compared with $T/R_b\simeq29.465$ for the truncated prescription at the same $M_0$ and $R_b/l_0$. Both times use clocks normalized at infinity, but the two models have different total masses at fixed bare-source parameters. The following exact lapse and delay coefficients refer to the first, explicitly truncated prescription.
\end{revision}

\begin{figure*}[tbp]
\includegraphics[width=\textwidth]{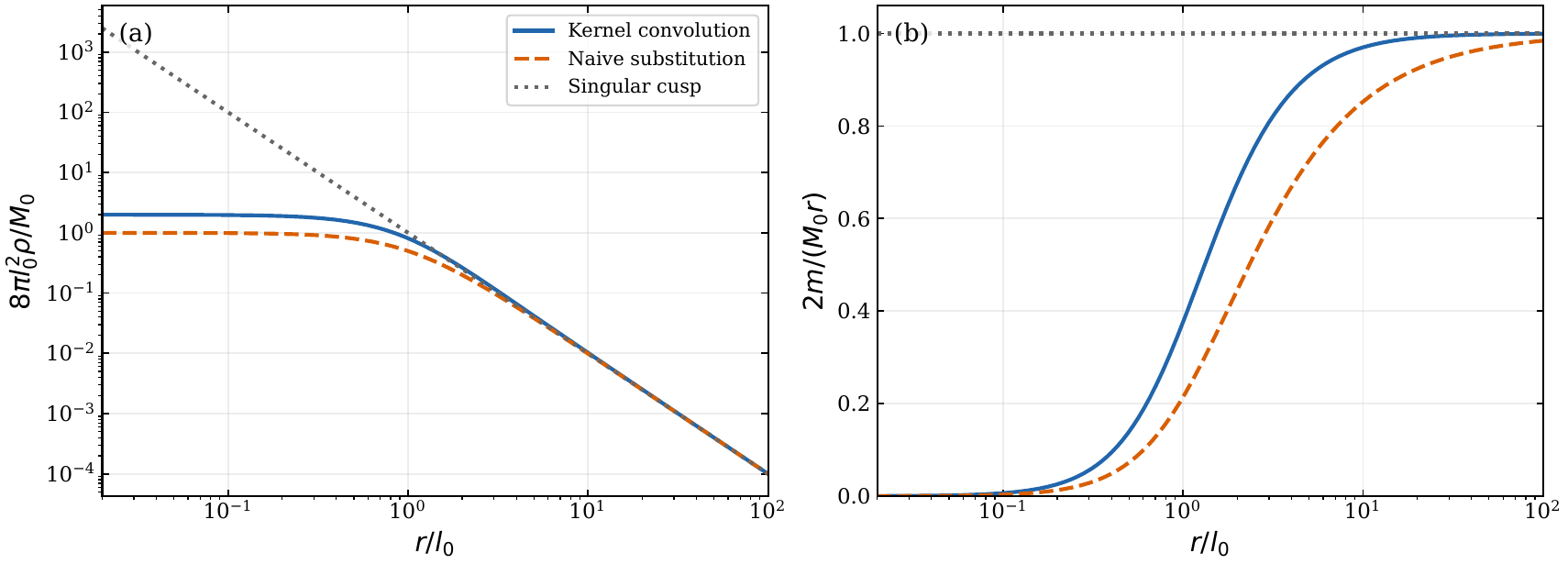}
\caption{\revised{Source profiles against $x=r/l_0$ for the auxiliary unbounded cusp. Panel (a) shows the dimensionless density $8\pi l_0^2\rho/M_0$: normalized kernel convolution, Eq.~\eqref{eq:rho} (solid blue), naive density substitution (dashed orange), and singular $x^{-2}$ cusp (dotted grey). Their regular central limits are 2 and 1, respectively, while all three approach $x^{-2}$ at large $x$. Panel (b) shows the normalized compactness $2m/(M_0r)$: $1-h(x)$ for convolution, $1-\arctan(x)/x$ for naive substitution, and 1 for the singular model. Both regular curves vanish at the centre and approach 1, with different mass-deficit asymptotics. The normalization removes dependence on $M_0$; the plotted interval is not a specified finite boundary.}}
\label{fig:profiles}
\end{figure*}

\subsection{Comparison with other prescriptions}

\textit{Naive substitution.} 
A common shortcut is to replace
$r^{2}\to r^{2}+\lz^{2}$ directly in the density. For JMN-1 this keeps only
the first term of Eq.~\eqref{eq:rho} and gives
\begin{equation}
m_{\rm naive}(r)=\frac{\Mz}{2}\left[r-\lz\arctan\frac{r}{\lz}\right].
\end{equation}
The centre is again regular, but the central density is half of the correct
value and the mass at large radius falls short of the JMN-1 value by
$\pi\Mz\lz/4$.
The shortcut does not arise from the normalized convolution. The second
term in Eq.~\eqref{eq:rho} supplies the doubled central density and removes
the constant large-radius mass deficit of the naive prescription
(Fig.~\ref{fig:profiles}). This is an asymptotic comparison for the
untruncated cusp, not a proof of finite-cloud mass conservation after
truncation at $\Rb$.

\textit{Gaussian smearing.} In models inspired by noncommutative geometry, the
delta function is replaced by a Gaussian of width
$\sqrt{2\theta}$~\cite{Nicolini:2006}, so that
$\tilde K=e^{-\theta k^{2}}$. The same steps, with
\begin{equation}
\int_{0}^{\infty}\sin(kr)\,e^{-\theta k^{2}}dk
=\theta^{-1/2}D(r/2\sqrt{\theta}),
\end{equation}
give
\begin{equation}
\rho_{G}(r)=\frac{\Mz}{8\pi}\,\frac{1}{r\sqrt{\theta}}\,
D\!\left(\frac{r}{2\sqrt{\theta}}\right),\,\,\,\rho_{G}(0)=\frac{\Mz}{16\pi\theta},
\end{equation}
where $D(x)=e^{-x^{2}}\int_{0}^{x}e^{t^{2}}dt$ is the Dawson function. The Gaussian parameter $\theta>0$ has dimension length squared. The
structure is the same: a constant-density core of size $\sqrt{\theta}$ that
joins the $1/r^{2}$ profile. The approach to JMN-1 is faster, with
corrections of relative order $\theta/r^{2}$.

\section{The regular geometry}
\label{sec:geometry}

\subsection{Field equations}

\revised{We impose the closure $p_r=0$ within the classical class of
static tangential-pressure interiors~\cite{Einstein:1939,Florides:1974},
and by analogy with JMN-1.} Density smearing alone does not determine the
pressure components, and a nonlocal stress-tensor modification need not
preserve zero radial pressure. Our choice specifies an effective static
matter model; its microscopic realization and dynamics are not derived
here. In particular, its ultracompact members cannot be massive-particle
Einstein clusters throughout their interiors. Equations~\eqref{eq:ee1}--\eqref{eq:ee3} then
reduce to
\begin{equation}
\begin{gathered}
1-\frac{2m}{r}=1-\Mz\,[1-h],\\
\Phi'=\frac{m}{r\,(r-2m)},\qquad
p_{\theta}=\frac{\rho\,m}{2\,(r-2m)},
\end{gathered}
\label{eq:field}
\end{equation}
with $h$ evaluated at $x=r/\lz$. In terms of $x$,
\begin{equation}
\begin{gathered}
\frac{d\Phi}{dx}=\frac{\Mz\,(1-h)}{2x\,[1-\Mz(1-h)]},\\
\frac{p_{\theta}}{\rho}=\frac{\Mz\,(1-h)}{4\,[1-\Mz(1-h)]}.
\end{gathered}
\label{eq:Phix}
\end{equation}
Equations~\eqref{eq:rho}, \eqref{eq:mass}, and \eqref{eq:field} define the
zero-point-length-corrected JMN-1 spacetime. For $\lz\to0$ at fixed $r$ one
has $h\to0$ and Eqs.~\eqref{eq:source} and \eqref{eq:jmn} are recovered. The
function $\Phi$ is given by one quadrature, which we evaluate numerically;
its limits are analytic.

\subsection{The core}

For $r\ll\lz$, Eq.~\eqref{eq:hsmall} gives
\begin{align}
m&\simeq\frac{\Mz r^{3}}{3\lz^{2}},&
\rho&\simeq\frac{\Mz}{4\pi\lz^{2}}\left(1-\frac{4r^{2}}{3\lz^{2}}\right),
\nonumber\\
\Phi&\simeq\Phi(0)+\frac{\Mz r^{2}}{6\lz^{2}},&
p_{\theta}&\simeq\frac{\Mz^{2}}{24\pi\lz^{2}}\,\frac{r^{2}}{\lz^{2}}.
\label{eq:core}
\end{align}
Hence $g_{rr}\to1$, the metric is smooth at the centre, and the fluid is
isotropic there, $p_{\theta}=p_{r}=0$, as regularity demands.
To leading order near the centre, the spatial metric agrees with a
constant-curvature three-sphere of radius $\lz\sqrt{3/(2\Mz)}$.
Higher-order terms in $m(r)$ prevent this identification from being exact
throughout the core. The curvature invariants are finite,
\begin{equation}
R(0)=\frac{2\Mz}{\lz^{2}},\qquad
R_{\mu\nu\rho\sigma}R^{\mu\nu\rho\sigma}(0)=\frac{20\,\Mz^{2}}{3\,\lz^{4}}.
\label{eq:curv}
\end{equation}
The divergence $\propto r^{-4}$ of Eq.~\eqref{eq:Kjmn} is cut off at
$r\sim\lz$.
The metric functions have regular even central expansions. In local
Cartesian coordinates the centre is a smooth point, and geodesics extend
through it; a radial continuation crosses to the opposite angular
direction rather than introducing a second asymptotic region. For each
fixed $\lz>0$ and finite $\Rb$, the lapse and $1-2m/r$ remain positive
on the entire interior. With the shell-free, asymptotically flat exterior
and the full static time range, causal geodesics encounter no finite-affine-
parameter boundary. The boundary has finite jumps in matter and curvature,
not a distributional thin shell. \latest{Section~\ref{sec:propertime} gives the continuation argument including nonradial causal geodesics.}

The core is not a de Sitter core. It has nearly constant density and
vanishing radial pressure; tangential stresses grow as $r^2$ away from
the centre and support equilibrium.
The central region admits an Einstein-cluster interpretation where the
circular speed is subluminal. This interpretation does not extend to the
whole interior of an ultracompact member.

\subsection{No horizon, no exotic matter}

Since $0\leq1-h<1$, one has
\begin{equation}
\frac{2m}{r}=\Mz\,[1-h(r/\lz)]<\Mz<1
\end{equation}
everywhere, so there is no horizon and no trapped surface, for any value of
$\lz$. The density \eqref{eq:rho} is positive. By Eq.~\eqref{eq:Phix},
$p_{\theta}/\rho$ rises monotonically from zero at the centre to its JMN-1
value $a/2$ at $r\gg\lz$. Hence
\begin{equation}
\rho>0,\qquad 0\leq p_{\theta}\leq\frac{a}{2}\,\rho,
\end{equation}

The weak and strong energy conditions therefore hold for $0<\Mz<1$.
For the finite object define
\begin{equation}
X\equiv\Rb/\lz,\qquad
C_b\equiv\frac{2m(\Rb)}{\Rb}=\Mz[1-h(X)].
\label{eq:Cb}
\end{equation}
We also write $C(r)\equiv2m(r)/r$ for the local compactness. Since $C(r)$
increases monotonically, the dominant energy
condition holds throughout the interior if and only if $C_b\le4/5$.
The familiar condition $\Mz\le4/5$ is sufficient and becomes the exact
limiting bound as $X\to\infty$. Energy conditions do not, by themselves,
establish a causal microscopic matter model or its perturbative stability.

There is no trapped region in this construction. Familiar regular black
holes with de Sitter-like cores violate the strong energy condition
\cite{Bardeen:1968,Hayward:2006,Ansoldi:2008}; this should not be
interpreted as a theorem following from central regularity alone without
additional global assumptions. Here the imposed closure produces a
regular centre while satisfying that condition.

\subsection{Exterior and total mass}

For our explicitly truncated density and $p_r=0$, the Darmois--Israel
conditions give a shell-free matching to Schwarzschild at $\Rb$.
The exterior mass and lapse normalization are exactly
\begin{equation}
M=m(\Rb)=\frac{\Mz\Rb}{2}[1-h(X)],\qquad
e^{2\Phi(\Rb)}=1-C_b.
\end{equation}
The relative mass correction to $\Mz\Rb/2$ is $h(X)=O(X^{-2}\ln X)$.
This finite model does not preserve exactly the mass of a bare cloud with
the same $\Mz$ and $\Rb$. For an actual convolution of that finite bare
cloud, the full integrated mass is preserved only if its exterior tail is
included. Discarding that tail loses, to leading order for $X\gg1$,
$\Delta M\sim\Mz\lz/4$, rather than the $O(X^{-2}\ln X)$ relative
correction of the present prescription.

At fixed $M$ the vacuum exterior is exactly Schwarzschild. An exterior
photon sphere, when present, has $r=3M$ and critical impact parameter
$b_c=3\sqrt3M$. Its existence and position do not alone imply an identical
observed shadow: brightness also depends on interior propagation,
emissivity, opacity, illumination history, and delayed returning light.

\subsection{Finite central redshift}

In JMN-1 the centre is an infinite-redshift point, $g_{tt}(0)=0$.

Let $I(x)=\int_0^x (d\Phi/du)\,du$. Shell-free matching fixes the exact
finite-boundary lapse as
\begin{equation}
e^{2\Phi(r)}=(1-C_b)\exp\{2[I(r/\lz)-I(X)]\}.
\label{eq:lapseexact}
\end{equation}
At fixed $\Mz$ and $X\gg1$, its central value has the asymptotic form
\begin{equation}
-g_{tt}(0)\simeq(1-\Mz)\,\kappa(\Mz)\left(\frac{\lz}{\Rb}\right)^{2a},
\label{eq:gtt0}
\end{equation}
with relative corrections $O(X^{-2}\ln X)$ at fixed $\Mz<1$. The number $\kappa$ is fixed by
a convergent integral,
\begin{equation}
\ln\kappa=-2\lim_{X\to\infty}\left[\int_{0}^{X}\!
\frac{\Mz\,(1-h)\,dx}{2x\,[1-\Mz(1-h)]}-a\ln X\right].
\label{eq:kappa}
\end{equation}
For small compactness it can be evaluated analytically,
\begin{equation}
\ln\kappa=(1-\ln2)\,\Mz+O(\Mz^{2}),
\end{equation}
and numerical values are listed in Table~\ref{tab:kappa}.
For $\Mz=1/2$, the integral can be evaluated exactly:
\begin{equation}
I(x)=\frac12\ln\left[
\frac{\sqrt{1+x^2}+\sinh^{-1}(x)/x}{2}\right].
\label{eq:Ihalf}
\end{equation}
Differentiation reproduces Eq.~\eqref{eq:Phix}, and the regular limit
fixes $I(0)=0$. Since $I(x)=\frac12\ln x-\frac12\ln2+o(1)$,
Eq.~\eqref{eq:kappa} proves $\kappa(1/2)=2$. Combining
Eq.~\eqref{eq:Ihalf} with the exact boundary normalization gives
\begin{equation}
-g_{tt}(0)=\frac{1}{\sqrt{1+X^2}}
=\frac{\lz}{\sqrt{\Rb^2+\lz^2}}
\simeq\frac{\lz}{\Rb}.
\end{equation}

\begin{table}[tbp]
\caption{The constant $\kappa(\Mz)$ of Eq.~\eqref{eq:gtt0}, the exponent
\revised{$a$, and the formal stable-light-ring radius of the auxiliary unbounded
profile, Eq.~\eqref{eq:lr}. A finite root exists for $\Mz>2/3$;
for the finite cloud it represents an interior light ring only when
$r_{\rm LR}<\Rb$, equivalently $C_b>2/3$. The $\infty$ entry denotes
the limiting formal radius, not a light ring of a finite cloud.}}
\label{tab:kappa}
\begin{ruledtabular}
\begin{tabular}{cccc}
$\Mz$ & $a$ & $\kappa$ & $r_{\rm LR}/\lz$ \\
\hline
0.10 & 0.056 & 1.041 & --- \\
0.20 & 0.125 & 1.114 & --- \\
0.30 & 0.214 & 1.241 & --- \\
0.40 & 0.333 & 1.483 & --- \\
0.50 & 0.500 & 2.000 & --- \\
0.60 & 0.750 & 3.416 & --- \\
2/3 & 1.000 & 6.326 & $\infty$ \\
0.68 & 1.063 & 7.46 & 12.9 \\
0.70 & 1.167 & 9.89 & 7.5 \\
0.72 & 1.286 & 13.8 & 5.7 \\
0.75 & 1.500 & 26.0 & 4.4 \\
0.80 & 2.000 & 128.5 & 3.3 \\
\end{tabular}
\end{ruledtabular}
\end{table}

\begin{figure*}[tbp]
\includegraphics[width=\textwidth]{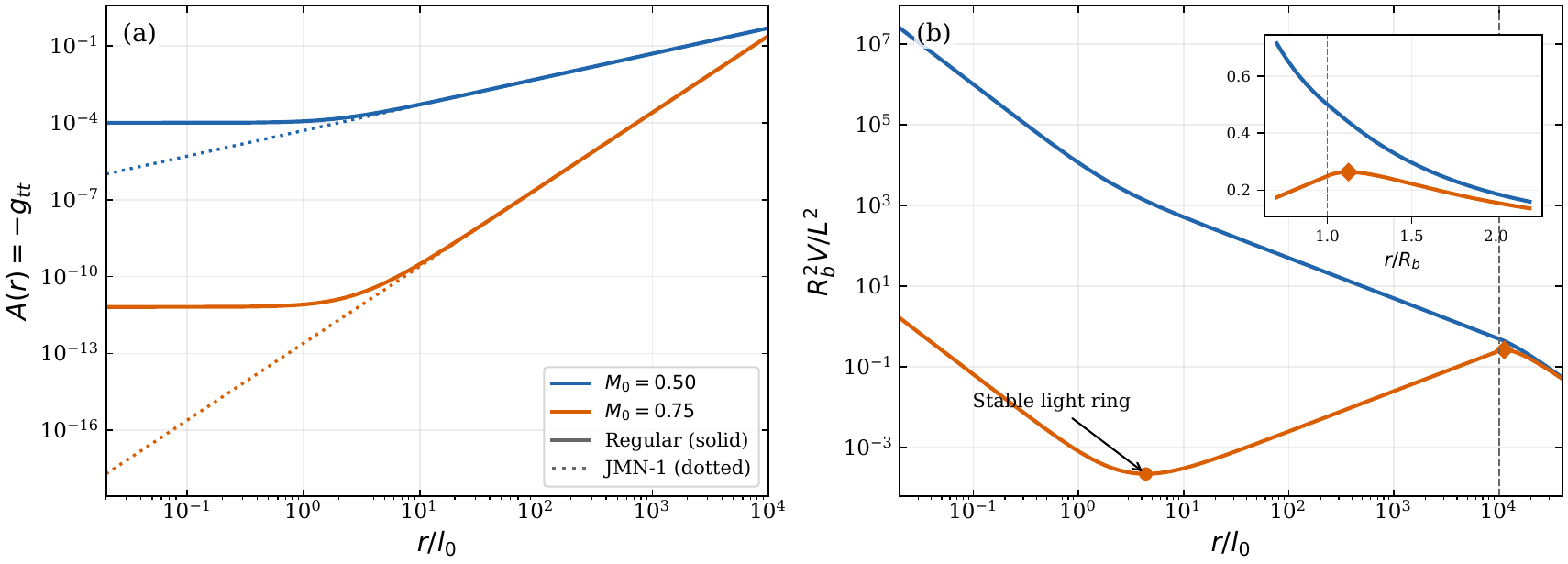}
\caption{\revised{Metric and null-orbit structure for the sharply truncated prescription with $X=R_b/l_0=10^4$, $M_0=0.50$ (blue), and $M_0=0.75$ (orange). Panel (a) plots the positive lapse $A(r)=-g_{tt}$ against $r/l_0$ for the regular interior (solid) and singular JMN-1 (dotted), each with its corresponding exterior normalization. Regular curves saturate at finite central values and recover the envelope away from the core. Panel (b) plots $R_b^2V/L^2=R_b^2A/r^2$, including each Schwarzschild exterior. The vertical dashed line marks $R_b$. For $M_0=0.75$, a circle identifies the stable interior minimum at $r/l_0=4.3754$ and a diamond the exterior maximum at $r=3M$; the inset resolves the latter against $r/R_b$. The $M_0=0.50$ potential decreases monotonically. The chosen $X$ displays both scales and does not represent a Planck-scale astrophysical cloud.}}
\label{fig:metric}
\end{figure*}

The redshift of a signal emitted by a static source at the centre and
received at infinity is
\begin{equation}
1+z_c=[-g_{tt}(0)]^{-1/2}
\simeq\frac{1}{\sqrt{(1-\Mz)\kappa}}\left(\frac{\Rb}{\lz}\right)^a.
\label{eq:zc}
\end{equation}
The zero-point length thus trades the singularity for a core of very large
but finite redshift. Figure~\ref{fig:metric}(a) shows $-g_{tt}$ for two
values of $\Mz$: the JMN-1 power law holds down to $r\sim\lz$ and then
saturates.

\section{Geodesics}
\label{sec:geodesics}

\subsection{Timelike circular orbits}

\revised{For equatorial geodesics we set $\vartheta=\pi/2$. Along a massive test particle, a dot denotes proper-time differentiation, $E=e^{2\Phi}\dot t$ and $L=r^2\dot\varphi$ are conserved energy and angular momentum per unit rest mass, and $\Omega=d\varphi/dt$ is the angular frequency measured with the asymptotically normalized time.} For the metric \eqref{eq:metric}, circular timelike geodesics have
\begin{equation}
E^{2}=\frac{e^{2\Phi}}{1-r\Phi'},\qquad
L^{2}=\frac{r^{3}\Phi'}{1-r\Phi'},\qquad
\Omega^{2}=\frac{e^{2\Phi}\Phi'}{r},
\end{equation}
and the orbital velocity measured by static observers is
$v^{2}=r\Phi'=m/(r-2m)$. They exist where $r\Phi'<1$, that is, where
$2m/r<2/3$.

In JMN-1, $v^{2}=a$ is constant down to the singularity and
$\Omega\propto r^{a-1}$, which diverges at the centre for $\Mz<2/3$. With a
zero-point length, Eq.~\eqref{eq:core} gives
\begin{equation}
v\simeq\sqrt{\frac{\Mz}{3}}\,\frac{r}{\lz},\qquad
\Omega\to\frac{e^{\Phi(0)}}{\lz}\sqrt{\frac{\Mz}{3}},
\end{equation}
\revised{so the circular-geodesic angular frequency approaches a constant and
the orbital velocity vanishes at the centre. This describes the orbital
family; it does not imply coherent bulk rotation of the spherical fluid.}
For $\Mz<2/3$, circular timelike orbits exist at every nonzero interior
radius. More generally they exist wherever $C(r)<2/3$. Their radial
stability follows analytically: with $q=r\Phi'=C/[2(1-C)]$, one has
$q'>0$, and
\begin{equation}
\frac{dL^2}{dr}
=\frac{2rq}{1-q}+\frac{r^2q'}{(1-q)^2}>0
\qquad(q<1).
\end{equation}
\revised{This establishes radial test-particle stability for circular orbits
strictly inside the cloud, not stability of the self-gravitating matter.
At the sharply matched surface the second derivatives of the potential
can jump, so stability of an orbit exactly at $\Rb$ must be assessed
from both sides. Exterior circular timelike orbits have the usual
Schwarzschild existence and stability bounds $r>3M$ and $r>6M$,
respectively, restricted to the vacuum region.} If $C_b>2/3$, interior timelike circular orbits
exist only below the stable light ring $r_{\rm LR}$. The central
angular-frequency divergence of singular JMN-1 is cut off; for Planck-scale
$\lz$ the directly modified disk region is microscopic.

For a massive-particle Einstein cluster, the kinetic stress relation is
$2p_\theta/\rho=v^2<1$. Our effective stress tensor obeys this restriction
only where $C<2/3$. For $C>2/3$ it requires $v^2>1$, so the outer portion
of an ultracompact member has no such kinetic realization. The large-delay
solutions below must therefore be interpreted as effective anisotropic
fluids whose microscopic support has not been specified.

\subsection{Light rings}

\revised{For null geodesics a dot instead denotes differentiation with respect to an affine parameter $\lambda$, and $E=e^{2\Phi}\dot t$ and $L=r^2\dot\varphi$ are the corresponding conserved quantities. Their separate magnitudes depend on affine normalization, but $b=L/E$ is the invariant impact parameter with dimension length.} Null geodesics with angular momentum $L$ move in the potential
\begin{equation}
V(r)=\frac{L^{2}e^{2\Phi}}{r^{2}}.
\end{equation}
In the JMN-1 interior $V\propto r^{2a-2}$. For $\Mz<2/3$ it decreases
monotonically and there is no light ring. For $\Mz>2/3$ it increases with
$r$ and falls to zero at the singularity; the only light ring is the unstable
photon sphere at $r=3M$ in the exterior, which exists because
$\Rb=2M/\Mz<3M$.

\revised{At the singular threshold $\Mz=2/3$, $a=1$ and $V$ is constant
throughout the JMN-1 interior. Circular null geodesics then form a
degenerate family at all nonzero interior radii, with no isolated minimum
or maximum of $V$.}

With a zero-point length the centrifugal barrier is restored at $r\ll\lz$,
where $V\simeq L^{2}e^{2\Phi(0)}/r^{2}$. For $\Mz>2/3$ the potential then has
a minimum where $r\Phi'=1$, that is, where $2m/r=2/3$:
\begin{equation}
h\!\left(\frac{r_{\rm LR}}{\lz}\right)=1-\frac{2}{3\Mz}.
\label{eq:lr}
\end{equation}

For the auxiliary unbounded profile, this equation has a unique finite
solution if and only if $\Mz>2/3$. For the actual finite object the ring
lies strictly inside the cloud if and only if $C_b>2/3$; otherwise the
formal root is at or beyond the matching radius. At the same boundary
compactness, the Schwarzschild photon sphere lies strictly outside the
cloud. The interior ring is stable because $C'(r)>0$ makes $r\Phi'-1$
increase through zero, changing $V'$ from negative to positive. Its
radius is of order a few to tens of $\lz$ for the parameters in
Table~\ref{tab:kappa}, but it diverges near the unbounded-profile
threshold. Close to the threshold
Eq.~\eqref{eq:hlarge} gives
$r_{\rm LR}^{2}/\lz^{2}\simeq\ln(2r_{\rm LR}/\lz)/[1-2/(3\Mz)]$, so the ring
moves outward and disappears as $\Mz\to2/3$.

At $\Mz=2/3$ and finite $X$, the regularized cloud instead has
$C_b<2/3$ and no circular null orbit. A distinct marginal case occurs
when $C_b=2/3$: then $\Rb=3M$ and the formal interior root and the
Schwarzschild circular null orbit meet at the matching surface. Since
$V'<0$ immediately on both sides, this boundary orbit is not a strict
local minimum or maximum. The pair of separated stable and unstable
rings exists only for $C_b>2/3$.

For $C_b>2/3$, the stable interior ring and unstable exterior photon
sphere form a pair consistent with the light-ring results for regular
horizonless ultracompact objects~\cite{Cunha:2017}. Their existence and
stability are established directly here, without requiring global
smoothness of the sharply truncated density at the matching surface. The singular JMN-1 solution evades that theorem; the
regularized one obeys it. Null rays with impact parameter $b$ such that
$V_{\rm min}<L^{2}/b^{2}<L^{2}/(27M^{2})$ and that start between the two rings are trapped between their radial turning points. Returning rays
that penetrate inside the exterior photon sphere can also generate
additional relativistic images, a characteristic feature of transparent
horizonless ultracompact objects~\cite{Shaikh:2019lensing}.

\subsection{Light-crossing time}

The coordinate time for a radial light ray to cross the cloud is
\begin{equation}
T=2\int_{0}^{\Rb}\frac{e^{-\Phi}\,dr}{\sqrt{1-2m/r}}.
\label{eq:Tdef}
\end{equation}

For fixed $a<1$ and $X\to\infty$, large radii dominate the integral and
its leading term approaches the JMN-1 value,
\begin{equation}
T\sim\frac{2\Rb}{(1-\Mz)(1-a)},\qquad \Mz<\tfrac{2}{3},
\end{equation}

The leading term is independent of $\lz$, but the exact finite-$X$
crossing time is not. This asymptotic limit is not uniform near $a=1$;
there the exact quadrature should be used. For $a\ge1$, the singular
JMN-1 crossing integral diverges, consistently with the central null
boundary described by its tortoise-coordinate limit. With a zero-point length the core
cuts the divergence off. Writing $e^{-\Phi(r)}=e^{-\Phi(0)}F(r/\lz)$ with
$F(x)=\exp[-\int_{0}^{x}(d\Phi/dx')dx']$ and using Eq.~\eqref{eq:gtt0},
\begin{equation}
T\simeq\tau(\Mz)\,\Rb\left(\frac{\Rb}{\lz}\right)^{a-1},
\qquad \Mz>\tfrac{2}{3},
\label{eq:T}
\end{equation}
with
\begin{equation}
\tau=\frac{2}{\sqrt{(1-\Mz)\kappa}}
\int_{0}^{\infty}\frac{F(x)\,dx}{\sqrt{1-\Mz[1-h(x)]}}.
\label{eq:tau}
\end{equation}
The integral converges because $F\simeq\sqrt{\kappa}\,x^{-a}$ at large $x$.

Numerically, $\tau=95.5413$, $35.0864$, $19.6391$, $10.0181$, and
$3.31163$ for $\Mz=0.68$, $0.70$, $0.72$, $0.75$, and $0.80$.
The slowly decaying tail must be retained. For a dimensionless cutoff
$Y\gg1$ it contributes to $\tau$, to leading order,
\begin{equation}
\Delta\tau(Y)\simeq\frac{2Y^{1-a}}{(1-\Mz)(a-1)}.
\end{equation}
This term is particularly important at $\Mz=0.68$, where $a-1=0.0625$.
For a large but finite cloud, extending the upper limit from $X$ to
infinity entails a relative error of order $X^{1-a}$; the asymptotic
\revised{formula is therefore also nonuniform at the threshold. At fixed $1<a<3$,
the first finite-boundary correction is
\begin{equation}
T=\tau\Rb X^{a-1}
-\frac{2\Rb}{(1-\Mz)(a-1)}+o(\Rb),
\qquad X\to\infty.
\label{eq:Tfinitecorrection}
\end{equation}
The remainder includes the lapse-normalization correction as well as
higher tail terms; this form applies for $1<a<3$, which includes the
DEC-satisfying large-$X$ range considered numerically here.
Near the threshold $\tau$}
diverges as $2/[(1-\Mz)(a-1)]$, and at $\Mz=2/3$ the power law is replaced
by a logarithm,
\begin{equation}
T\simeq6\,\Rb\ln\frac{\Rb}{\lz},\qquad \Mz=\tfrac{2}{3}.
\label{eq:Tlog}
\end{equation}

\subsection{\revised{A uniform optical-delay crossover}}
\begin{figure*}[tbp]
\includegraphics[width=\textwidth]{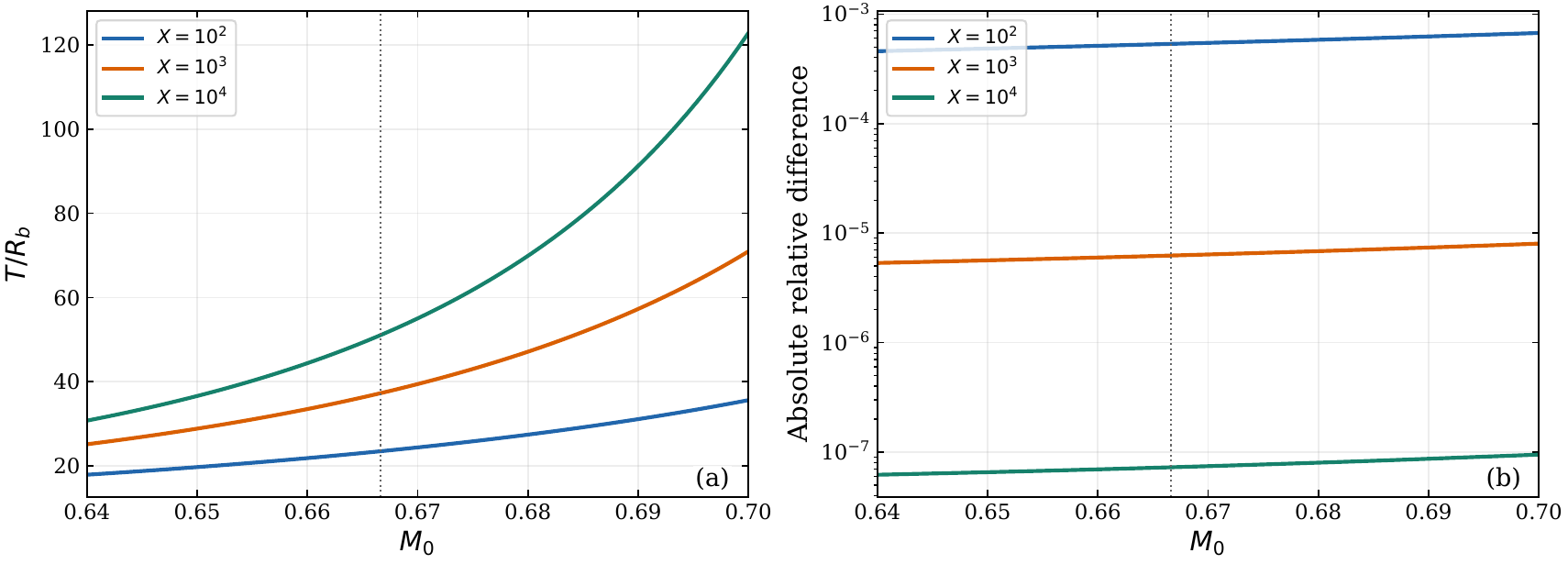}
\caption{\revised{Uniform transition of the dimensionless radial crossing time $T/R_b$ near $M_0=2/3$, for the sharply matched prescription. Panel (a) compares exact finite-$X$ quadrature (solid) with Eq.~\eqref{eq:uniformT} (dashed) for $X=10^2,10^3,10^4$; the vertical dotted line marks the critical parameter. The smooth crossover depends on $(a-1)\ln X$, rather than on a discontinuous change of geometry. Panel (b) plots the relative difference between the uniform expression and exact quadrature, with the same colour coding. The remainder decreases with $X$; its estimate is given in Eq.~\eqref{eq:uniformR}. The constant $\mathcal D(M_0)$ is evaluated from the convergent subtracted integral in Eq.~\eqref{eq:regularD}.}}
\label{fig:uniform}
\end{figure*}

\begin{revision}
The separate power laws can be joined without a divergent coefficient at the threshold. Define $\beta=1-M_0>0$, $\epsilon=a-1$, and
\begin{equation}
 f(x)=\frac{F(x)}{\sqrt{1-M_0[1-h(x)]}},
 \qquad d=\sqrt{\kappa/\beta}.
\end{equation}
Here $f$, $d$, $\beta$, and $\epsilon$ are dimensionless. The large-$x$ expansion of the lapse integral gives
\begin{equation}
 f(x)=d\,x^{-a}\left[1+O(x^{-2}\ln x)\right].
\end{equation}
Introduce a finite subtracted core integral and coefficient
\begin{align}
 H(M_0)&=\int_0^1 f(x)\,dx
       +\int_1^\infty[f(x)-d x^{-a}]\,dx,\\
 \mathcal D(M_0)&=\frac{2H(M_0)}{\sqrt{\beta\kappa}}.
 \label{eq:regularD}
\end{align}
The second integral converges for the physical $a>0$, and $\mathcal D$ is smooth in a neighbourhood of $a=1$. We use $\mathcal D$ to distinguish this coefficient from the Dawson function $D(x)$ used earlier. To avoid using divergent coefficients numerically, evaluate the explicitly subtracted integrand or its asymptotic tail.

For $X\gg1$, the exact integral satisfies
\begin{equation}
 \int_0^X f(x)\,dx
 =H+d\,\frac{1-X^{-\epsilon}}{\epsilon}
 +O(X^{-a-1}\ln X).
\end{equation}
Combining it with the exact boundary normalization gives
\begin{equation}
 \frac{T}{R_b}=\mathcal D X^\epsilon
 +\frac{2}{\beta}\frac{X^\epsilon-1}{\epsilon}
 +\mathcal R(X,M_0),
 \label{eq:uniformT}
\end{equation}
where the quotient means $\ln X$ at $\epsilon=0$. The remainder obeys, locally uniformly around $a=1$,
\begin{align}
 \mathcal R=O\biggl(X^{-2}\ln X
 \biggl[X^\epsilon
  +\left|\frac{X^\epsilon-1}{\epsilon}\right|+1\biggr]\biggr).
 \label{eq:uniformR}
\end{align}
For any fixed $0<a<3$, it is $o(1)$. This range includes the dominant-energy-condition envelope of the model. The expansion is not asserted uniformly as $M_0\to1$. The transition variable is $\epsilon\ln X$: it resolves the narrow compactness interval where the power law changes to a logarithm.

For $a<1$, Eq.~\eqref{eq:uniformT} yields the original finite JMN-1 limit. For $a>1$, it gives both the leading coefficient and first finite-boundary correction,
\begin{equation}
 \tau=\mathcal D+\frac{2}{\beta(a-1)}.
\end{equation}
At the threshold, independent quadrature gives
\begin{equation}
 \frac{T}{R_b}=6\ln X-4.1569938+o(1),
 \qquad M_0=\frac23.
 \label{eq:thresholdconstant}
\end{equation}
The finite constant improves on Eq.~\eqref{eq:Tlog} and explains its small difference from finite-$X$ propagation. Figure~\ref{fig:uniform} compares Eq.~\eqref{eq:uniformT} with the exact radial quadrature across the transition. These are optical statements about the specified metric, not yet predictions of a quantum propagation theory.
\end{revision}

\revised{Equation~\eqref{eq:Tdef} defines radial propagation through the transparent}
cloud, from one boundary point through the centre to the opposite boundary.
Its coordinate time is normalized to the proper time of a static observer
at infinity. For $\Mz>2/3$ and $X\gg1$ it scales as $\lz^{1-a}$.
It is not the crossing time of every ray entering the photon sphere.
For nonzero angular momentum the radial equation is
\begin{equation}
\dot r^2=\left(1-\frac{2m}{r}\right)
\left(E^2e^{-2\Phi}-\frac{L^2}{r^2}\right).
\end{equation}
Such rays generally turn before the centre. \revised{For an ultracompact matched object with $C_b>2/3$, the unstable exterior photon sphere lies at $r=3M>R_b$. Rays incident from infinity with $b=L/E<b_c$ can enter the interior and later re-emerge, whereas rays approaching $b_c$ can acquire additional arbitrarily long delays near this exterior photon sphere. That exterior trapping mechanism is absent when $C_b<2/3$; the marginal surface orbit at $C_b=2/3$ was distinguished above.} Thus a radiative-transfer calculation must use the full
impact-parameter-dependent travel times.

\subsection{\latest{Proper time of a radially infalling observer}}
\label{sec:propertime}
\begin{added}
The crossing time $T$ of Eq.~\eqref{eq:Tdef} is a boundary-to-boundary
null coordinate time, normalized to the proper time of static observers
at infinity. The proper time of a freely falling test observer is a
different quantity. Write $A(r)\equiv e^{2\Phi(r)}$ and
$B(r)\equiv1-2m(r)/r$ for the dimensionless positive metric functions.
For a future-directed radial timelike geodesic of Eq.~\eqref{eq:metric},
the conserved energy per unit rest mass is
$E=A\,dt/d\tau_{\rm p}>0$, where $\tau_{\rm p}$ is proper time and is
unrelated to the dimensionless optical coefficient $\tau(\Mz)$ of
Eq.~\eqref{eq:tau}. Four-velocity normalization gives
\begin{equation}
 \left(\frac{dr}{d\tau_{\rm p}}\right)^2
 =B\left(\frac{E^2}{A}-1\right).
 \label{eq:radialtimelike}
\end{equation}
Here $E$ is dimensionless and $\tau_{\rm p}$ has dimension length in
geometrized units. Below, $\tau_{\rm p}$ also denotes the elapsed proper
time of the inward boundary-to-centre segment.

In singular JMN-1, $A=(1-\Mz)(r/\Rb)^{2a}$ and $B=1-\Mz$.
For an inward geodesic starting at the boundary, with
$E^2\geq1-\Mz$, the elapsed proper time is
\begingroup\color{black}
\begin{equation}
 \tau_{\rm p}=\int_0^{\Rb}
 \frac{(r/\Rb)^a\,dr}
 {\sqrt{E^2-(1-\Mz)(r/\Rb)^{2a}}}.
 \label{eq:taudef}
\end{equation}
\endgroup
Its integrand vanishes as $r^a$ at the centre, whereas the null
coordinate-time integrand in Eq.~\eqref{eq:Tdef} grows as $r^{-a}$.
The boundary endpoint is integrable even for release from rest.
Thus $\tau_{\rm p}$ is finite for every fixed $0<\Mz<1$, including
$\Mz\geq2/3$, where the singular coordinate-time integral diverges.
For release from rest, $E^2=1-\Mz$; substituting
$y=(r/\Rb)^{2a}$ gives
\begingroup\color{black}
\begin{equation}
 \tau_{\rm p}^{\rm JMN}
 =\frac{\sqrt\pi\,\Rb}{2a\sqrt{1-\Mz}}
 \frac{\Gamma\!\left(\frac{1+a}{2a}\right)}
 {\Gamma\!\left(\frac{1+2a}{2a}\right)},
 \label{eq:tauclosed}
\end{equation}
\endgroup
where $\Gamma$ is the Euler gamma function. This yields
$\tau_{\rm p}^{\rm JMN}/\Rb=2.221441$, $\sqrt3=1.732051$, and
$1.493668$ for $\Mz=1/2$, $2/3$, and $3/4$, respectively.
The infaller reaches the singularity in finite proper time, although
its limiting static coordinate time is infinite for $\Mz\geq2/3$.
The clock mismatch originates in $dt/d\tau_{\rm p}=E/A$.
Finite infaller time and divergent static time also occur at a
Schwarzschild horizon, but that horizon is regular; the JMN-1 centre
is a curvature singularity.

For the sharply matched regularized object, release from rest at
$r=\Rb$ instead fixes $E^2=A(\Rb)=1-C_b$. The exact elapsed proper
time follows from
\begin{equation}
 \tau_{\rm p}=\int_0^{\Rb}
 \frac{\exp[I(r/\lz)-I(X)]\,dr}
 {\sqrt{B(r)}\sqrt{1-\exp\{2[I(r/\lz)-I(X)]\}}},
 \label{eq:regularpropertime}
\end{equation}
using the lapse of Eq.~\eqref{eq:lapseexact}. Direct quadrature gives
$\tau_{\rm p}/\Rb=1.734182912$, $1.732089359$, and $1.732051411$
at $\Mz=2/3$, for $X=10^2$, $10^3$, and $10^4$, respectively.
At $\Mz=3/4$ the corresponding values are $1.495837138$,
$1.493702279$, and $1.493668863$. At fixed $\Mz=3/4$, the difference
from Eq.~\eqref{eq:tauclosed} is $O(X^{-2}\ln X)$, but the threshold
has an additional logarithm:
\begin{equation}
 \frac{\tau_{\rm p}-\tau_{\rm p}^{\rm JMN}}{\Rb}
 =\frac{\sqrt3}{4}\frac{(\ln X)^2}{X^2}
 +O\!\left(\frac{\ln X}{X^2}\right),
 \quad \Mz=\frac23.
 \label{eq:properthreshold}
\end{equation}
To see its origin, at the threshold $B=(1+2h)/3$ and the
large-$x$ lapse integral obeys
\begin{equation}
 I(x)=\ln x-\frac12\ln\kappa
 +\frac{3}{2x^2}\left[\ln(2x)+\frac12\right]
 +O\!\left(\frac{(\ln x)^2}{x^4}\right).
 \label{eq:properIthreshold}
\end{equation}
Consequently, $e^{I(x)}/\sqrt{B(x)}$ contains a correction
proportional to $(\ln x)/x$. Integrating it across the envelope in
Eq.~\eqref{eq:regularpropertime} gives the squared logarithm;
boundary normalization and the turning endpoint contribute only at
$O(X^{-2}\ln X)$. Both corrections are negligible at astrophysical
scale ratios, despite the strong core sensitivity of $T$.

For the adopted Sgr~A$^*$ mass, the boundary-to-centre proper times
are approximately $110.05$~s at $\Mz=2/3$ and $84.36$~s at $\Mz=3/4$.
The corresponding boundary-to-opposite-boundary proper times are
$2\tau_{\rm p}\simeq220.10$~s and $168.72$~s. These specify the same
spatial endpoints as the null crossing time, for which
Table~\ref{tab:numbers} gives approximately $11$~h and
$5.8\times10^{17}$~yr in the asymptotically normalized clock.
The short timelike proper time does not imply a short coordinate time.
Indeed, define $t_{\rm fall}$ as the elapsed coordinate time from the
boundary to the centre. Equation~\eqref{eq:radialtimelike} gives
\begin{equation}
 t_{\rm fall}=\int_0^{\Rb}
 \frac{e^{-\Phi}\,dr}
 {\sqrt B\sqrt{1-A/E^2}}
 >\frac{T}{2}.
 \label{eq:coordinatefall}
\end{equation}
For $\lz>0$ this integral is finite; the square-root singularity at
the release point is integrable. By symmetry, the timelike
boundary-to-boundary coordinate time is $2t_{\rm fall}>T$.

The decisive geometric change is the endpoint. In singular JMN-1 a
radial timelike geodesic reaches $r=0$ at the finite proper time above.
For a radial null geodesic, $E=A\,dt/d\lambda>0$ instead denotes the
affine-normalization-dependent Killing energy, with $\lambda$ an affine
parameter. Its equation $|dr/d\lambda|=E(\Rb/r)^a$ gives the finite
boundary-to-centre interval
\begin{equation}
 \Delta\lambda=\frac{\Rb}{(a+1)E}.
 \label{eq:propernullaffine}
\end{equation}
The Kretschmann scalar of Eq.~\eqref{eq:Kjmn} diverges along these
geodesics. This excludes a regular $C^2$ extension through their central
endpoint, without asserting inextendibility in every weaker regularity
class. Both timelike and null geodesic incompleteness therefore occur
for every $0<\Mz<1$, on both sides of $\Mz=2/3$.

For each fixed $\lz>0$, finite $\Rb$, and $0<\Mz<1$, the regularized
centre instead has $A(0)>0$ and $B(0)=1$. The released radial observer
has $E^2>A(0)$ and crosses the centre with finite nonzero velocity in
regular local Cartesian coordinates. Continuation follows the opposite
angular direction, rather than another asymptotic region. Since the
nonnegative areal radius has a cusp at this crossing, its derivative
there is understood one-sidedly. The observer reaches the opposite
turning point $r=\Rb$ after a total proper time $2\tau_{\rm p}$.
Return to the same spatial position and velocity takes
$4\tau_{\rm p}$, whereas the areal radius alone repeats after
$2\tau_{\rm p}$.

Completeness also requires control of nonradial geodesics. Let
$g^{(3)}_{ij}$ be the metric on a static spatial slice,
\begin{equation}
 g^{(3)}_{ij}\,dy^i dy^j=B^{-1}dr^2+r^2d\Omega^2,
 \label{eq:properspatialmetric}
\end{equation}
where $y^i$ are local spatial coordinates and $i,j=1,2,3$.
For any future-directed causal geodesic, choose $\lambda$ to be proper
time for a timelike trajectory and an affine parameter for a null
trajectory. Define $\delta=1$ and $0$, respectively, so that
$g_{\mu\nu}(dx^\mu/d\lambda)(dx^\nu/d\lambda)=-\delta$.
Conservation of $E=A\,dt/d\lambda$ yields the full spatial and temporal
velocity bounds
\begin{align}
 g^{(3)}_{ij}\frac{dy^i}{d\lambda}\frac{dy^j}{d\lambda}
 &=\frac{E^2}{A}-\delta
 \leq\frac{E^2}{A(0)}=E^2(1+z_c)^2,
 \label{eq:speedbound}\\
 0<\frac{dt}{d\lambda}
 &=\frac{E}{A}\leq\frac{E}{A(0)}.
 \label{eq:propertimebound}
\end{align}
The lapse has its global minimum at the centre because $\Phi'\geq0$
in the interior and the Schwarzschild exterior. The spatial geometry
is complete: its centre is regular, it has no finite-distance boundary,
and its asymptotically flat end lies at infinite proper distance.
The bounds above keep both position and velocity within a compact
region of the local phase space on any finite affine interval, so
local geodesic evolution can be continued. At the shell-free matching
surface, Gaussian normal coordinates make the metric $C^{1,1}$,
meaning continuously differentiable with locally Lipschitz first
derivatives; the finite matter jump therefore does not obstruct
geodesic continuation. With the full static time range
$t\in\mathbb R$, the matched spacetime is causally geodesically complete
in both time directions. This direct argument is consistent with
standard static-spacetime completeness criteria~\cite{Sanchez:2005}.
It also applies to the smooth fully convolved finite cloud, whose lapse
in Eq.~\eqref{eq:finitelapse} has a positive minimum and whose spatial
end is complete. In singular JMN-1 the centre is absent from the
spatial manifold and $A^{-1}$ diverges there. Completeness here is
a property of the specified global effective metrics, rather than a
general consequence of inserting a zero-point length. 

The present construction differs from zero-point-length regular black-hole models with a de Sitter-like core, where short-distance gravity can become repulsive~\cite{Jusufi:2024dtr}. Repulsion alone, however, does not establish global geodesic completeness, which requires control of the full spacetime and its geodesic extensions. Here the core is not de Sitter, as noted in Sec.~\ref{sec:geometry}: the central density is positive, the pressures vanish at the centre, and $\Phi'(r)>0$ for $r>0$, so the gravitational field remains attractive and vanishes smoothly at the centre. A radial test observer released from rest at the boundary encounters no inner turning point; its speed relative to static observers increases inward, and it passes through the regular centre. As established above, causal geodesic completeness follows from the regular centre, the strictly positive minimum lapse, the complete spatial slices, and the extendibility of geodesics across the matching surface. No repulsive region or violation of the strong energy condition is required.
Finally, the radial observer's Lorentz factor relative to the static
observer at the centre is
\begingroup\color{black}
\begin{equation}
 \gamma_c=Ee^{-\Phi(0)}=\sqrt{1-C_b}\,(1+z_c).
 \label{eq:propergamma}
\end{equation}
\endgroup
It is finite for fixed $\lz>0$ but can be enormous at astrophysical
scale ratios. All trajectory statements assume a test particle that
does not interact with the cloud. As for the optical delay, the
Planckian core curvature and extreme local energy limit their physical
interpretation; causal completeness of the effective metric does not
establish a microscopic quantum-gravity description or dynamical
stability of its source.
\end{added}

\subsection{\revised{Illustrative astrophysical scales}}
\label{sec:numbers}

We now evaluate Eqs.~\eqref{eq:zc} and \eqref{eq:T} with $\lz$ equal to the
Planck length, $1.6\times10^{-35}$~m. For Sgr~A$^*$ we take
\revised{$M=4.3\times10^6M_\odot$, consistent with stellar-orbit astrometry
\cite{GRAVITY:2022}; for M87$^*$ we adopt
$M=6.5\times10^9M_\odot$~\cite{EHT:2019VI}. The
$10M_\odot$ case is an illustrative stellar-mass scale, rather than a
fit to a specific source.}
We use $\Rb\simeq2GM/(c^2\Mz)$; the exact relation is
$M=\Mz\Rb[1-h(X)]/2$ in geometrized units. Its correction is negligible
for these numbers. We restore $G$ and $c$ in numerical conversions;
$\Rb/\lz$ ranges from about $10^{39}$ to $10^{48}$.

\begin{table}[tbp]
\caption{Central redshift for the mass of Sgr~A$^*$ and light-crossing time
for three masses, with $\lz$ equal to the Planck length. For $\Mz=0.50$ the
crossing time is the leading large-$X$ JMN-1 limit; for $\Mz=2/3$ it is the
leading logarithm of Eq.~\eqref{eq:Tlog}. \revised{For $\Mz>2/3$, the entries use the leading asymptotic expression
\eqref{eq:T}; they are not exact finite-$X$ quadrature values.}}
\label{tab:numbers}
\begin{ruledtabular}
\begin{tabular}{ccccc}
 & $\log_{10}z_{c}$ & \multicolumn{3}{c}{$T$} \\
$\Mz$ & Sgr~A$^*$ & $10\,M_{\odot}$ & Sgr~A$^*$ & M87$^*$ \\
\hline
0.50 & 22.6 & 1.6 ms & 11 min & 12 d \\
2/3 & 44.9 & 81 ms & 11 h & 2.0 yr \\
0.68 & 47.7 & 4.03 s & 45.2 d & 295 yr \\
0.70 & 52.3 & 5.10 h & $2.17\times10^{3}$ yr & $1.11\times10^{7}$ yr \\
0.72 & 57.6 & 15 yr & $2.7\times10^{8}$ yr & $3.3\times10^{12}$ yr \\
0.75 & 67.1 & $2.1\times10^{9}$ yr & $5.8\times10^{17}$ yr & $3.4\times10^{22}$ yr \\
0.80 & 89.3 & $3.0\times10^{28}$ yr & $5.5\times10^{39}$ yr & $1.3\times10^{46}$ yr \\
\end{tabular}
\end{ruledtabular}
\end{table}

\begin{figure*}[tbp]
\includegraphics[width=\textwidth]{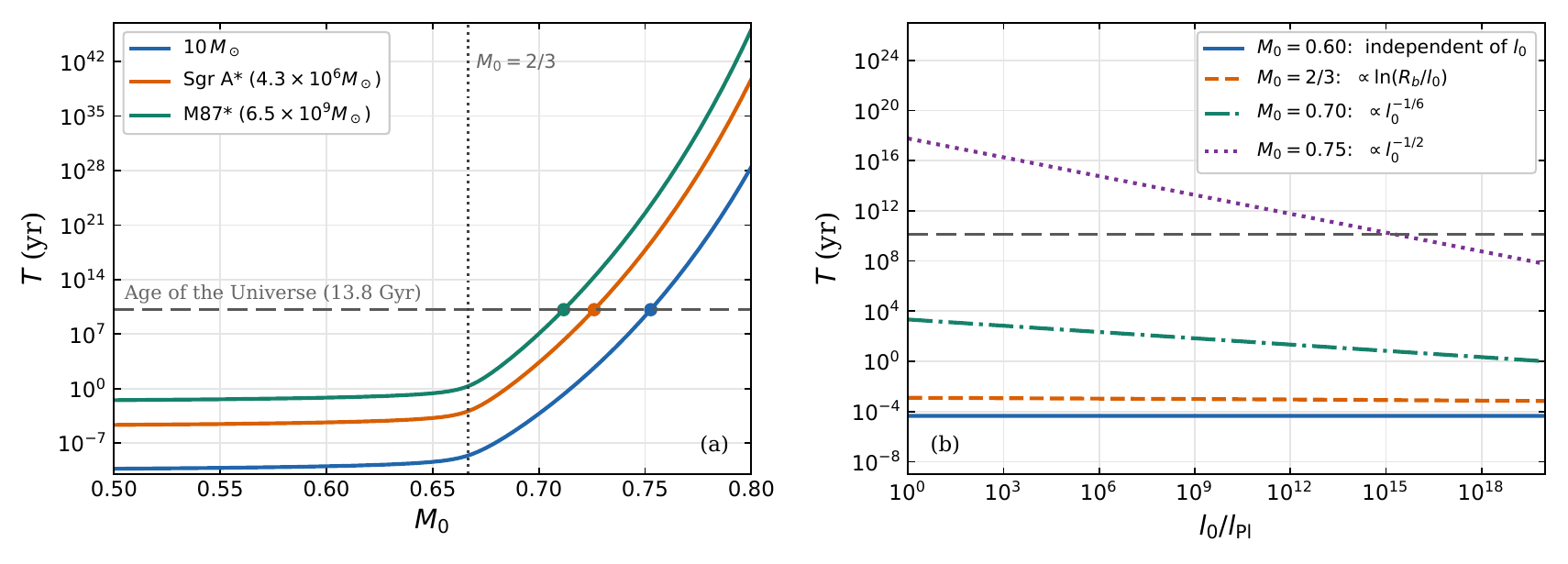}
\caption{\revised{Radial boundary-to-boundary crossing time, Eq.~\eqref{eq:Tdef}, in years, with the Schwarzschild time normalized at infinity and the sharply matched prescription; $R_b\simeq2GM/(c^2M_0)$ and the exact mass correction is negligible at these scale ratios. Panel (a): $T$ against $M_0$ for $10M_\odot$, Sgr~A$^*$ with $M=4.3\times10^6M_\odot$, and M87$^*$ with $M=6.5\times10^9M_\odot$, using $l_0=1.6\times10^{-35}$ m. Curves use the dimensionless core quadrature plus its asymptotic tail to the finite upper limit, remaining continuous through the dotted $M_0=2/3$ line; circles mark the three intersections with the age of the Universe. Panel (b): $T$ against $l_0/l_{\rm Pl}$, with $l_{\rm Pl}=1.6\times10^{-35}$ m, for the Sgr~A$^*$ mass at fixed $M_0$. Below the threshold $T$ is independent of $l_0$ at leading order, at $M_0=2/3$ it depends on $l_0$ only logarithmically, and above it $T\propto l_0^{1-a}$. The dashed horizontal line in both panels is 13.8 Gyr. The uniform formula explains the crossover, while Table~\ref{tab:numbers} reports the separate leading asymptotic values. These are formal transparent-metric propagation times; Planckian curvature and extreme local photon blueshift limit their direct observational interpretation.}}
\label{fig:T}
\end{figure*}

\revised{The small distinction between Table~\ref{tab:numbers} and the finite-$X$
curves in Fig.~\ref{fig:T} is controlled by the asymptotic corrections.
At $\Mz=0.68$, the leading formula gives $T=4.0343$~s for
$10M_\odot$ and $T=45.1661$~d for Sgr~A$^*$; retaining the finite
upper limit gives $4.0198$~s and $45.0940$~d, respectively.
The table deliberately reports the leading asymptotic values.
These estimates and Fig.~\ref{fig:T} exhibit three regimes.}

(i) For $\Mz<2/3$ the crossing time is a few light-crossing times of the
cloud, with no dependence on $\lz$ at leading order in $X\gg1$. The zero-point length shows up
only in the central redshift, which is finite but enormous: $10^{22.6}$ for
$\Mz=1/2$ and the mass of Sgr~A$^*$.

(ii) In the window $2/3<\Mz\lesssim0.73$ the delay is long but finite, and
it is extremely sensitive to both $\Mz$ and $\lz$. For Sgr~A$^*$ it grows
from hours to the age of the Universe over this window. If $\lz$ exceeds the
Planck length by a factor $10^{n}$, the delay is reduced by a factor
$10^{n(a-1)}$.

(iii) For larger $\Mz$ the radial delay exceeds the age of the Universe,
\revised{taken here as $13.8\times10^9$~yr, the rounded age inferred within
base $\Lambda$CDM from the Planck analysis~\cite{Planck:2020}.} The crossover values are
$\Mz\simeq0.75267$, $0.72605$, and $0.71169$ for the three masses in
Table~\ref{tab:numbers}. Delayed central signals can then be effectively
unobservable on available timescales, despite the absence of a horizon.
The exterior photon-sphere scale agrees with Schwarzschild, but a
transparent regular interior does not permanently capture an open set of
incident null rays. An apparent dark image region depends on illumination
history, opacity, and emission, and is not guaranteed to be identical to
the Schwarzschild shadow by the metric alone. The singular JMN-1 imaging
results~\cite{Shaikh:2019} cannot be transferred without these additional
assumptions.

\begin{revision}
These numbers assume transparency and classical propagation through a core of size $l_0$. The assumption can be tested against the local photon energy,
\begin{equation}
 E_{\rm loc}(r)=h_{\rm P}\nu_\infty e^{-\Phi(r)},
 \qquad E_{\rm Pl}=\sqrt{\frac{\hbar c^5}{G}},
 \label{eq:localphoton}
\end{equation}
where $h_{\rm P}=2\pi\hbar$ is Planck's constant and $\nu_\infty$ is the frequency measured at infinity. We distinguish $h_{\rm P}$ from the mass-profile function $h(x)$. For a hypothetical 230-GHz photon reaching the centre of the Sgr~A$^*$ example at $M_0=0.68$, $1+z_c\simeq10^{47.7}$ gives $E_{\rm loc}(0)\simeq5\times10^{44}$ eV, approximately $4\times10^{16}E_{\rm Pl}$. Its local wavelength would be about $3\times10^{-51}$ m, well below $l_0$. Thus the large delays are reliable integrals of the prescribed metric but are not controlled predictions for actual millimetre photons traversing a Planck-scale core. At Planckian curvature there is also no broad wavelength interval simultaneously larger than the microscopic cutoff and smaller than the curvature scale. For a larger effective $l_0$, both curvature and the local-energy hierarchy must be reassessed; the delay alone does not establish validity.
\end{revision}

\section{\revised{Scalar probes and pressure-closure dependence}}
\label{sec:probe}

\subsection{\revised{A nonnegative massless scalar operator}}
\begin{revision}
The static source does not determine coupled fluid or gravitational perturbations, but a minimally coupled massless scalar probe has a well-defined equation on either prescribed background. Denote its field by $\chi$, satisfying $\Box\chi=0$, where $\Box$ is the metric wave operator. Write
\begin{equation}
 \chi=\frac{u_\ell(t,r)}{r}Y_{\ell j}(\vartheta,\varphi),
 \qquad A=e^{2\Phi},\quad B=1-\frac{2m}{r}.
\end{equation}
$Y_{\ell j}$ is a spherical harmonic, $\ell=0,1,\ldots$ its angular index, and $j=-\ell,\ldots,\ell$ its azimuthal index, distinguished from the mass function $m$. The radial field $u_\ell$ can be complex; physical real fields are recovered by conjugate combinations. Define the tortoise coordinate by $dr_*/dr=(AB)^{-1/2}$ and set $r_*(0)=0$. Its range is $[0,\infty)$ for either regular model. Separation gives
\begin{align}
 (\partial_t^2+\mathcal H_\ell)u_\ell&=0,\\
 \mathcal H_\ell&=-\partial_{r_*}^2+V_\ell,\nonumber\\
 V_\ell&=A\frac{\ell(\ell+1)}{r^2}+\frac{(AB)'}{2r}.
 \label{eq:scalarV}
\end{align}
The potential has dimension length$^{-2}$. Regularity selects $u_\ell=O(r^{\ell+1})$ at the centre. In the sharply matched model $u_\ell$ and $\partial_{r_*}u_\ell$ are continuous at $R_b$; $A$ and $B$ are continuous, so a finite jump of $V_\ell$ introduces no delta function.

Let $f=\sqrt{AB}$ and $\mathcal Q=\partial_{r_*}-f/r$. The factorization
\begin{equation}
 \mathcal H_\ell=\mathcal Q^\dagger\mathcal Q
                  +A\frac{\ell(\ell+1)}{r^2}
 \label{eq:factorization}
\end{equation}
follows from $\partial_{r_*}=f\partial_r$ and
$\partial_{r_*}(f/r)=ff'/r-f^2/r^2$. It is an application of the operator positivity technique underlying $S$-deformations~\cite{Ishibashi:2003}, with $S=-f/r$. For regular compactly supported radial data, boundary terms vanish at the centre and infinity and cancel at the matching surface. Therefore
\begin{equation}
 \langle u,\mathcal H_\ell u\rangle
 =\int_0^\infty\left[|\mathcal Q u|^2
    +A\frac{\ell(\ell+1)}{r^2}|u|^2\right]dr_*\geq0.
 \label{eq:probeenergy}
\end{equation}
The factorization is a general identity for minimally coupled massless scalar probes on regular static spherical backgrounds with positive $A$ and $B$; its model-specific application here includes the centre and matching conditions and the explicit trapping potential. The regular finite-energy self-adjoint realization inherits this inequality. In particular it has no negative eigenvalue $\omega^2$ that would produce an exponentially growing normalizable mode. Equivalently, the quadratic wave energy
\begin{equation}
 \mathcal E_\ell=\frac12\int_0^\infty
 [|\partial_tu_\ell|^2+|\mathcal Q u_\ell|^2
  +A\ell(\ell+1)|u_\ell|^2/r^2]dr_*
\end{equation}
is nonnegative and conserved when no energy escapes the domain. This is scalar probe mode stability, not a proof of decay or stability of the effective matter. Indeed, the $\ell=0$ potential can be negative near the centre: the core expansion gives $V_0(0)=-A(0)M_0/(3l_0^2)$. Positivity follows from the factorization, rather than from positivity of the potential pointwise.
\end{revision}

\subsection{\revised{Trapped-mode scales and semiclassical leakage}}
\begin{figure*}[tbp]
\includegraphics[width=\textwidth]{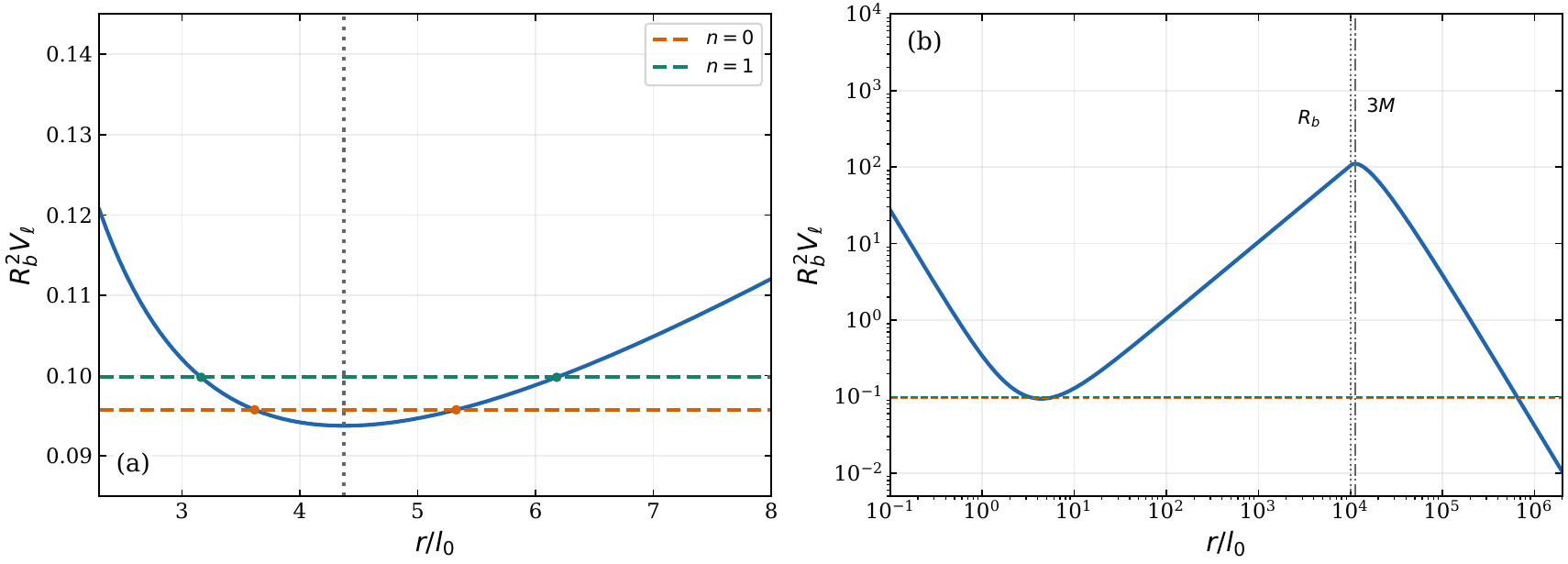}
\caption{\revised{Massless scalar potential and leading WKB well estimates for the sharply matched model, $M_0=0.75$, $X=10^4$, $\ell=20$. Panel (a) plots $R_b^2V_\ell$ over core radii $r/l_0$, with horizontal dashed lines at $(\omega_RR_b)^2$ for $n=0,1$. Their intersections identify the two well turning radii used in Eq.~\eqref{eq:WKBwell}; the vertical dotted line identifies the stable null light ring, which is close to but not exactly the scalar-potential minimum. Panel (b) shows the same potential through the exterior barrier on logarithmic axes, marking $R_b$ and $3M$. The inset-free broad radial range includes the exterior turning points. The positive barrier actions in Table~\ref{tab:wkb} suppress scalar tunnelling. Scalar operator nonnegativity is proved separately and does not imply rapid decay or matter stability.}}
\label{fig:scalar}
\end{figure*}

\begin{revision}
For large $\ell$, $V_\ell$ is dominated by $\ell(\ell+1)A/r^2$, so the stable light ring produces a potential well separated from infinity by the outer barrier~\cite{Cardoso:2014,Keir:2016}. Use $u_\ell\propto e^{-i\omega t}$, where $\omega$ is the complex frequency in the time normalized at infinity; outgoing resonances have $\operatorname{Im}\omega<0$. In a deep well with three simple turning radii $r_1<r_2<r_3$, the leading Wentzel--Kramers--Brillouin (WKB) prescription is
\begin{align}
 \int_{r_1}^{r_2}\sqrt{\omega_R^2-V_\ell}\,dr_*
      &=\pi(n+\tfrac12),\label{eq:WKBwell}\\
 S_\ell&=\int_{r_2}^{r_3}\sqrt{V_\ell-\omega_R^2}\,dr_*.
 \label{eq:WKBbarrier}
\end{align}
Here $\omega_R>0$ approximates the real part, $n=0,1,\ldots$ labels the radial overtone, and $S_\ell$ is the dimensionless barrier action. The leading tunnelling probability is $\mathcal P_\ell\sim e^{-2S_\ell}$. Its prefactor and conversion to a damping rate require the well oscillation time and higher-order matching. We report the barrier factor, rather than an uncomputed exact imaginary frequency.

For the illustrative matched geometry $M_0=0.75$, $X=10^4$, Table~\ref{tab:wkb} gives leading WKB values, and Fig.~\ref{fig:scalar} displays the well and barrier. The exterior potential used in this calculation is
\begin{equation}
 V_\ell^{\rm ext}=(1-2M/r)
       \left[\frac{\ell(\ell+1)}{r^2}+\frac{2M}{r^3}\right].
\end{equation}
The finite surface jump of the subleading scalar term is included piecewise; the eikonal term is continuous. WKB accuracy improves with angular index in a deep well, but no exact complex eigenfrequency or rigorous WKB error bound is claimed. The large barrier actions diagnose suppressed scalar leakage. They do not establish a lifetime for material or gravitational perturbations, nor prove a nonlinear instability.

Nonlinear scalar evolutions on other fixed backgrounds with stable trapping exhibit energy cascades even where the linear problem has bounded, slowly decaying solutions~\cite{Benomio:2025}. Those results motivate a separate nonlinear probe study here; they do not establish instability or collapse of the present effective source.

Define the angular frequency of the stable light ring by $\Omega_s=e^{\Phi(r_{\rm LR})}/r_{\rm LR}$ and its dimensionless radius by $x_s=r_{\rm LR}/l_0$. At fixed $M_0>2/3$ and large $X$,
\begin{align}
 \Omega_s&\simeq\frac{\sqrt{\beta\kappa}\,e^{I(x_s)}}{x_sR_b}
                      X^{1-a},\\
 \Omega_s T&\longrightarrow
       \frac{\tau\sqrt{\beta\kappa}\,e^{I(x_s)}}{x_s}.
 \label{eq:modeTime}
\end{align}
For fixed radial overtone and large angular index, the leading real trapped-mode frequency satisfies $\omega_R/\sqrt{\ell(\ell+1)}\to\Omega_s$. Thus a long radial optical delay and a small asymptotic trapped-mode frequency share the same lapse scale. Their product depends on the dimensionless core structure. This supplies a concrete bridge from the static redshift calculation to a controlled probe spectrum, while a gravitational-wave interpretation still requires the matter dynamics.
\end{revision}

\begin{table}[tbp]
\caption{\revised{Leading WKB scalar well frequencies and barrier transmission factors for $M_0=0.75$, $X=10^4$. The frequency column is the dimensionless product $\omega_RR_b$, not $\omega_R$ in physical units. $S_\ell$ is Eq.~\eqref{eq:WKBbarrier}; the final column is $-2S_\ell/\ln10$, the logarithm of the leading tunnelling factor. These are semiclassical estimates, not exact quasinormal frequencies or fluid lifetimes.}}
\label{tab:wkb}
\begin{ruledtabular}
\begin{tabular}{ccccc}
$\ell$ & $n$ & $\omega_RR_b$ & $S_\ell$ & $\log_{10}\mathcal P_\ell$\\
\hline
20 & 0 & 0.309436 & 361.298 & -313.820 \\
20 & 1 & 0.315907 & 358.007 & -310.961 \\
40 & 0 & 0.608157 & 715.566 & -621.533 \\
40 & 1 & 0.614629 & 711.928 & -618.373 \\
\end{tabular}
\end{ruledtabular}
\end{table}

\subsection{\revised{What changes when radial pressure is introduced}}
\begin{revision}
To separate source smoothing from the pressure closure, keep the same density and mass but prescribe $p_r=w(r)\rho$, where the dimensionless function $w(r)$ is smooth, even at the centre, and vanishes at $R_b$ in the sharply matched model. The field equations then determine
\begin{align}
 \Phi_w'&=\frac{m+4\pi r^3w\rho}{r(r-2m)},\\
 p_{\theta,w}&=w\rho+\frac r2(w\rho)'
       +\frac r2(1+w)\rho\Phi_w'.
 \label{eq:closurew}
\end{align}
With finite $w(0)$, the centre remains regular and $p_{\theta,w}(0)=p_r(0)$; imposing $w(R_b)=0$ retains shell-free Schwarzschild matching. The function $w$ is an equilibrium prescription, not an equation for dynamical perturbations or a specified sound speed. Energy conditions must be checked using the full stresses, especially across the taper where $w'$ contributes.

Let $\Phi_0$ denote the original $p_r=0$ lapse potential. Both closures have the same mass and boundary value $\Phi_w(R_b)=\Phi_0(R_b)=\tfrac12\ln(1-C_b)$. Subtracting their radial field equations and integrating gives the exact relation
\begin{equation}
\begin{split}
 \Phi_w(r)-\Phi_0(r)
 &=-\int_r^{R_b}
 \frac{4\pi s^2 w(s)\rho(s)}{s-2m(s)}\,ds .
\end{split}
\label{eq:pressurelapse}
\end{equation}
For $w(r)\geq0$ throughout the interior, the integrand is nonnegative, so $\Phi_w\leq\Phi_0$. Consequently the central redshift cannot decrease and the radial crossing time $T_w=2\int_0^{R_b}e^{-\Phi_w}[1-2m/r]^{-1/2}dr$ obeys $T_w\geq T$. This comparison is exact at fixed density, mass, and boundary radius; it does not establish the energy conditions or stability of the pressure-modified source.

In a broad JMN-like envelope $l_0\ll r\ll R_b$ where $w(r)=w_e$ is constant, $0\leq w_e\leq1$, and $\rho\simeq M_0/(8\pi r^2)$,
\begin{align}
 r\Phi_w'&\simeq a_w=\frac{M_0(1+w_e)}{2(1-M_0)},\\
 \frac{p_{\theta,w}}\rho&\simeq
       \frac{M_0(1+w_e)^2}{4(1-M_0)}.
 \label{eq:wevelope}
\end{align}
For the scaling comparison, hold $M_0$ and $w_e$ fixed as $X=R_b/l_0\to\infty$. The regular core has fixed size $l_0$ and a fixed bounded pressure profile in $r/l_0$; the boundary taper has fixed fractional width and a fixed bounded profile in $r/R_b$. These profiles join the constant-$w_e$ envelope and introduce no additional $X$ dependence. Equation~\eqref{eq:pressurelapse} then contributes $-a w_e\ln X+O(1)$ to the central lapse potential, giving $e^{\Phi_w(0)}$ proportional to $X^{-a_w}$ with a finite nonzero matching coefficient. Envelope matching gives the same three delay regimes with $a$ replaced by $a_w$: $T_w/R_b$ tends to a finite constant for $a_w<1$, grows logarithmically for $a_w=1$, and is proportional to $X^{a_w-1}$ for $a_w>1$. The finite, logarithmic, and power-law regimes are separated by
\begin{equation}
 a_w=1\quad\Longleftrightarrow\quad
 M_0=\frac{2}{3+w_e}.
 \label{eq:wthreshold}
\end{equation}
The lapse coefficients depend on the chosen core and taper. For $0\leq w_e\leq1$, the envelope dominant condition additionally requires $M_0\leq4/[4+(1+w_e)^2]$, but this envelope inequality alone does not control the transition region. Since the mass is unchanged, horizon avoidance survives any regular closure with a positive finite lapse, whereas the enormous-delay exponent is closure dependent. At $w_e=0$ the envelope expressions recover those of the tangential-pressure model; the complete original geometry is recovered when $w(r)\equiv0$. This explicit comparison makes clear why a static density cannot uniquely predict optical delays without specifying its relativistic stresses.
\end{revision}

\section{\revised{Discussion and conclusions}}
\label{sec:discussion}

A normalized zero-point-length kernel smooths the integrable JMN-1 cusp and, with a specified tangential-pressure closure, produces a regular static geometry. The exactly truncated prescription has a Schwarzschild exterior without a thin shell; the fully convolved finite cloud instead retains its smooth matter tail and its original total mass. The finite-source compactness bound establishes that horizon avoidance is not an artefact of truncating the auxiliary infinite profile. For both zero-radial-pressure constructions the weak and strong energy conditions hold, and the dominant condition is governed by the maximum local compactness. The smooth centre is an ordinary point of a one-ended geometry, rather than a wormhole throat.

\begin{added}
Finite infaller proper time and long optical delay are compatible because
these observables use different clocks. The explicit radial infall
calculation retains a finite JMN-1 proper-time limit, with a squared
logarithmic correction at the critical compactness. The regular core
changes geodesic endpoints: the singular metric admits incomplete radial
causal geodesics, whereas the two regularized global metrics are causally
geodesically complete. This conclusion uses their complete spatial slices,
positive minimum lapse, and full static time range. It does not follow
from density smearing or bounded curvature alone, and it does not supply
a dynamical theory of singularity resolution.
\end{added}

The central lapse is finite and can be extremely small. The exact half-compactness solution verifies the numerical normalization, while the subtracted optical integral gives a uniform delay formula through the critical parameter. Its transition variable combines the envelope exponent with the scale hierarchy; the finite threshold constant completes the leading logarithmic result. These quantitative statements follow from the particular kernel and pressure closure. Smoothing with other regular kernels can reproduce the qualitative envelope-to-core transition, but changes its coefficients. Introducing radial pressure changes the delay exponent itself, as the explicit envelope calculation shows.

Ultracompact matched members possess a stable core light ring and an unstable exterior photon sphere. This pair is consistent with the established light-ring structure of regular horizonless compact objects~\cite{Cunha:2017}; it is not a new general light-ring theorem. The massless scalar probe has a nonnegative radial operator, excluding growing normalizable modes even though its zero-angular-index potential is locally negative. Semiclassical well modes exhibit strong barrier suppression and share the lapse scale underlying the optical delay. Nonnegative mode energy does not ensure rapid decay: stable trapping motivates concern about nonlinear behaviour~\cite{Cardoso:2014,Keir:2016}. Selected bosonic-star evolutions demonstrate light-ring-driven migration or collapse~\cite{Cunha:2023}, but do not determine the response of the present effective fluid.

The contribution beyond classical tangential-pressure interiors is therefore the explicit cusp convolution, its two consistent finite-source realizations, the finite-source compactness estimate, the controlled delay crossover, and the joint analysis of optical and scalar-probe scales. Regularity with zero radial pressure has classical precedents~\cite{Florides:1974}; energy-condition preservation alone is not a microscopic matter construction. In the ultracompact zero-radial-pressure region, the required circular speed exceeds the massive-particle Einstein-cluster bound. A causal material theory for that region, and its coupled perturbation equations, remain necessary.

These geometries also differ from metric-level Simpson--Visser regularizations~\cite{Simpson:2019,Pal:2023}, which turn the JMN construction into a wormhole supported by throat null-energy-condition violation. Familiar Bardeen and Hayward black holes have horizons and de Sitter-like cores~\cite{Bardeen:1968,Hayward:2006,Ansoldi:2008}. Here no trapped region is present, so preservation of the strong condition is compatible with regularity and the absence of the additional hypotheses of singularity theorems. The proposed thermodynamic-topology protection of regular black holes~\cite{Anand:2026} is a separate framework; our static effective source does not prove a dynamical statement about cosmic censorship.

Wormhole constructions motivated by the same T-duality regularization~\cite{LoboRodrigues:2025} provide a further comparison: smoothing the source does not by itself select a throat or its supporting stresses. The present pressure closure instead yields an ordinary regular centre and preserves the weak and strong energy conditions. The source prescription, relativistic stresses, and global topology must therefore be specified together.

At fixed measured mass the exterior of the first prescription is Schwarzschild, but its photon-sphere scale alone does not fix an observed dark region. Singular JMN-1 imaging calculations~\cite{Shaikh:2019} require new transport assumptions for a transparent regular centre. The illustrative Planck-scale delays extrapolate the metric through Planckian curvature and, for millimetre photons reaching the core, trans-Planckian local energies. Their interpretation consequently requires a propagation law and a matter-opacity model. The fully convolved finite source additionally modifies the exterior through its retained tail. Neither the formal radial time nor a scalar tunnelling factor can be identified directly with gravitational-wave echo spacing or damping~\cite{Cardoso:2016,Cardoso:2019}.

Future work can proceed along connected, testable steps. A covariant effective action or causal material closure would fix perturbation dynamics and allow the coupled stability problem to be posed. Exact scalar resonances and time-domain evolution would test the semiclassical estimates and quantify slow decay. Impact-parameter-dependent transport, including delayed returning light~\cite{Shaikh:2019lensing}, would determine whether finite-age illumination produces a dark central region and how it differs from a black-hole image. The smooth finite-cloud model provides a way to assess boundary-prescription dependence without discarding mass. Extensions to the second JMN solution and generalized families~\cite{Trivedi:2026} require independent convolution and closure checks. Finally, collapse, accretion, and rotation must be studied within a specified dynamical theory before these static regular configurations can be interpreted as astrophysical end states.

\begin{acknowledgments}
	FSNL acknowledges support from the Funda\c{c}\~ao para a Ci\^encia e a Tecnologia (FCT) research grant UID/04434/2025.
\end{acknowledgments}

\bibliographystyle{apsrev4-2}
\bibliography{refs}

\end{document}